\documentclass[aps,prd,superscriptaddress,showpacs,tighten,nofootinbib]{revtex4-1}
\usepackage{epsfig}
\usepackage{amssymb}
\usepackage{rotate}
\usepackage{float}
\usepackage{graphicx}
\usepackage{subfig}
\usepackage{slashed}
\usepackage{amsmath}
\usepackage{setspace}
\usepackage{slashed}
\usepackage{color}
\usepackage{comment}
\usepackage{dcolumn}   
\usepackage{bm}        
\usepackage{multirow}
\newcommand{\ba}{\begin{array}}
\newcommand{\ea}{\end{array}}
\newcommand{\bd}{\begin{displaymath}}
\newcommand{\ed}{\end{displaymath}}
\newcommand{\be}{\begin{equation}}
\newcommand{\ee}{\end{equation}}
\newcommand{\bea}{\begin{eqnarray}}
\newcommand{\eea}{\end{eqnarray}}

\def\a{\alpha}
\def\b{\beta}
\def\g{\gamma}

\def\ve{\varepsilon}

\def\m{\mu}
\def\n{\nu}

\def\n{\nu}

\def\th13 {\theta_{13}}

\usepackage{graphicx}

\newcommand{\mathsym}[1]{{}}
\newcommand{\unicode}[1]{{}}

\catcode`\@=11
\def\lsim{\mathrel{\mathpalette\@versim<}}
\def\gsim{\mathrel{\mathpalette\@versim>}}
\def\@versim#1#2{\vcenter{\offinterlineskip
\ialign{$\m@th#1\hfil##\hfil$\crcr#2\crcr\sim\crcr } }}
\catcode`\@=12

\catcode`@=12
\begin{document}
\title{ The Discovery reach of $CP$ violation in neutrino oscillation   with non-standard interaction effects}

\author{Zini Rahman}
\email{zini@ctp-jamia.res.in}
\affiliation{Centre for Theoretical Physics, Jamia Millia Islamia (Central University), \\ Jamia Nagar, New Delhi-110025, India}

\author{Arnab Dasgupta}
\email{arnab@ctp-jamia.res.in}
\affiliation{Centre for Theoretical Physics, Jamia Millia Islamia (Central University), \\ Jamia Nagar, New Delhi-110025, India}

\author{Rathin Adhikari}
\email{rathin@ctp-jamia.res.in}
\affiliation{Centre for Theoretical Physics, Jamia Millia Islamia (Central University), \\ Jamia Nagar, New Delhi-110025, India}

\begin{abstract}
We have studied the $CP$ violation discovery reach in neutrino oscillation experiment  with superbeam, neutrino factory 
and monoenergetic neutrino beam from electron capture process. For NSI satisfying  model-dependent bound for shorter baselines
(like CERN-Fr\'ejus set-up ) there is
insignificant  effect of NSI on the the discovery reach of $CP$ violation due to $\delta$. Particularly, for superbeam and neutrino factory
we have also considered relatively longer baselines for which there could be significant NSI effects on $CP$ violation discovery reach for higher allowed values of NSI. For
monoenergetic beam only shorter baselines are considered to study $CP$ violation with different nuclei as neutrino sources.
 Interestingly for  non-standard interactions - $\ve_{e\mu}$
and $\ve_{e\tau}$  of neutrinos with matter during propagation in longer baselines in superbeam,  there is 
possibility of better discovery reach of $CP$ violation than  that with only Standard Model interactions of neutrinos with matter. For complex NSI we have shown the $CP$ violation discovery reach in the plane of Dirac phase $\delta$ and
NSI phase $\phi_{ij}$. The $CP$ violation due to some values of $\delta $  remain unobservable  with present and near future experimental facilities
in superbeam and neutrino factory . However, in presence of some ranges of off-diagonal NSI phase values there 
are some possibilities of discovering total $CP$ violation for any $\delta_{CP}$ value even at $5 \sigma$ confidence level for neutrino factory. Our analysis indicates that for some values of  NSI phases total $CP$ violation may not be at all observable for any values of $\delta $. Combination of shorter and longer baselines could indicate in some cases the presence of NSI. However,
in general for NSIs $\lesssim 1$ the $CP$ violation discovery reach is better in neutrino factory set-ups.  Using a neutrino beam from electron capture process for nuclei $^{110}_{50}$Sn and $^{152}$Yb,  we have shown the discovery reach of $CP$ violation in neutrino oscillation experiment.  Particularly for $^{110}_{50}$Sn nuclei $CP$  violation could be found for about 51\%  of the possible $\delta$ values for a baseline of 130 km with boost factor $\gamma = 500$. The nuclei  $^{152}$Yb is although  technically more feasible for the production of mono-energetic beam, but  is found to be not  suitable in obtaining good discovery reach of $CP$ violation.

\end{abstract}
\maketitle
\section{Introduction}
Long back $CP$ violation  has been found in the quark sector of the Standard Model of Particle Physics. But so far there is no evidence of $CP$ violation in the leptonic sector. One way to search for such $CP$ violation is through neutrino oscillation experiments in which one flavor of neutrino could oscillate to other flavors of neutrinos. Neutrino oscillation probability depends on various oscillation parameters present in the $3 \times 3$ mixing matrix - the PMNS matrix \cite{pmns,bilen,part} and the neutrino mass squared differences. Two of the three angles - $\theta_{12}$ and $\theta_{23}$ present in PMNS matrix have been known with certain accuracy for some time. Recently  several experiments like Double CHOOZ, Daya Bay, RENO and T2K collaboration \cite{double,daya,reno,t2k} found non-zero value of $\sin^2 2 \theta_{13}$ corresponding to the third mixing angle of PMNS matrix with a global level of significance which is well above conventional  $5\sigma$ discovery threshold. As three  angles are non-zero  there could be
non-zero $CP$ violating phase $\delta $ in the PMNS matrix and as such there could be $CP$ violation in the leptonic sector.  The mass squared differences - $|\Delta m^2_{31}|$ and $\Delta m^2_{21}$ is known (where $\Delta m^2_{ij} = m_i^2 -m_j^2$) but the sign of $\Delta m^2_{31}$ and as such the hierarchy (whether it is  normal (NH) or inverted (IH)) of neutrino masses is still unknown.  Various neutrino oscillation experiments like superbeam, neutrino factory, beta beams and reactor experiments are focussing on determining these unknown parameters corresponding to neutrino oscillations \cite{lbne,lbno,beta,Edgecock:2011ova,others,others5,Coloma:2012ma,others4,Itow,Choubey,Zucc,Bouch}. 

There could be various kind of non-standard interactions of neutrinos with matter. This could have some effect on
finding $CP$ violation in neutrino oscillation experiments. Such interactions could be the non-standard
 interactions of neutrinos with matter fermions ($u, d$ and $e$) during propagation of neutrinos only. This could affect oscillations 
of different flavors of neutrinos as sub-leading effect. We have discussed it in further detail in the next section.   There could be 
other different kind of interactions beyond Standard model leading to  non-unitarity of
$3\times 3$ PMNS neutrino mixing matrix \cite{nu}.     Considering  non-standard interactions of neutrinos at the source and the detector in
 neutrino oscillation experiments also lead to such possibility. However, such NSI at the source and detector have highly stringent 
constraints \cite{Ohlsson} and as such the effect on neutrino oscillation is expected to be lesser affected than that due to NSI
in matter during propagation of neutrinos. We shall not consider NSI at the source and detector in this work.  

To study $CP$ violation in neutrino oscillation experiments we have considered superbeam, neutrino factory and monoenergetic 
neutrino beam from electron capture process. There are some earlier studies for  short and long baselines for
standard interactions \cite{raut,Dighe,Coloma:2012ma,Agarwalla:2011hh,others4,Coloma:2012ut,betabeam}
 for neutrinos coming from superbeam. We consider neutrino superbeam (which mainly contains $\nu_{\mu}$ and ${\bar \nu_{\mu}}$ ) coming from CERN travelling a baseline length of 2300 km  to Pyh\"asalmi (Finland) and another baseline of 130 km to  Fr\'ejus (France). We have also considered superbeam coming
from Tokai to Hyper-Kamiokande  travelling a baseline of 295 km. 
We do a comparative study in the discovery potentials of the $CP$ violating phase $\delta$ for these three baselines in presence of both standard and non-standard interactions. Main emphasis of our work is on exploring
$CP$ violation due to $\delta$ in the light of recent experimental findings of large value of $\theta_{13}$ and how NSI could affect the discovery reach of such $CP$ violation.  Also there are some studies on $CP$ violation
in presence of non-standard interactions \cite{nsi08p,Ota} before Daya Bay experiment when $\theta_{13}$ was not so precisely known.

The discovery of $CP$ violation for different parent muon energy in neutrino factory and different length of baselines 
have been 
studied earlier \cite{Ballett:2012rz,Agarwalla:2010hk}. In the  analysis done in 
\cite{Ballett:2012rz}  three different kinds of magnetized detectors: a Totally-Active Scintillator(TASD) and
two kinds of Liquid Argon detectors have been considered for baselines ranging from 1000 km to 4000 km and parent muon energies from 4 GeV to 25 GeV, whereas 
in the analysis done in \cite{Agarwalla:2010hk} MIND (Toroidal 
magnetized iron neutrino detector) detector has been considered. 
As recent reactor neutrino experiments indicates 
large value of $\theta_{13}$ it is important to study the discovery potential of $CP$ violation in neutrino oscillation 
experiments in low energy neutrino factory \cite{lownufact2}.  
There are some studies on the performance of low energy neutrino factories \cite{Choubey} in the context of standard model 
interactions of neutrinos with matter
\cite{Ballett:2012rz,lownufact2,lownufact1,pas2,pas3,pas4,pas5,pas6,pas7} and also in presence of NSI \cite{lowenergy3,nsi3,fried,Coloma:2011rq,nsi4}.
In the works related to NSI \cite{lowenergy3,nsi3,fried,Coloma:2011rq,nsi4} longer baselines like 4000 and 7500 kms has been considered to have larger NSI effect on neutrino oscillation experiments 
so as to find better NSI sensitivity and in that context $CP$ violation has also been studied. 

Our work is complementary to these earlier analysis in the sense that we focus here on the better discovery reach of $CP$ violation 
and as such we consider relatively shorter baselines where the matter effect and as such NSI effect will be lesser in 
the neutrino oscillation experiments. However, in the context of such experimental set-up in neutrino factory which is better
 optimized for $CP$ violation discovery (considering only SM interactions of neutrinos with matter) we have also addressed the 
question of  $CP$ violation coming due to NSI phases.  
 We have considered lower parent muon energy within 10 GeV which is found to be more favourable for the discovery of 
$CP$ violation. The optimization for $CP$ violation discovery in the context of SM interactions has been done earlier \cite{Ballett:2012rz,Agarwalla:2010hk}. However, in this work we re-analysed this optimization with the updated detector 
characteristics for the MIND detector \cite{mind1}. Based on this optimization 
we have chosen a few relatively shorter baselines (730, 1290 and 1500 kms) and low parent muon energy of 10 GeV.

In recent years oscillation experiment using a neutrino beam with neutrinos emitted from an electron capture process is proposed \cite{Zucc,rolinec,mon,mono2,mono3,mono9,mono11,mono12}. Such  beam can be produced using an accelerated nuclei that decay by electron capture. In this process an electron is captured by a proton releasing a neutron and an electron neutrino. So
the beam is purely of one flavor. In the rest frame of the mother nuclei the electron neutrino that is released from such process, has a definite energy Q. Since the idea of using a neutrino beam emitted from an electron capture process is based on the acceleration and storage of radioactive isotopes that decays to daughter nuclei, one may get the suitable neutrino energy by accelerating the mother nuclei with suitable Lorentz boost factor $\gamma$. One can control the neutrino energy by choosing the appropriate Lorentz boost factor  as the energy that has been boosted by an  appropriate boost factor towards the detector is given as $E=2 \gamma Q$. Hence for certain mother nuclei to get the required neutrino energy the  boost factor have to be chosen appropriately with respect to Q.  Due to the almost monoenergetic nature of  such beam  one can appropriately choose the neutrino energy for which the probability of oscillation could be large and sensitive to certain unknown  neutrino oscillation parameters.

In this work we consider such a flavor pure electron neutrino beam emitted from electron capture process for a suitable $\gamma$ value where the beam is targeted towards a Water Cherenkov detector and perform numerical simulation to study the discovery reach of $CP$ violation in oscillation experiments.

In section \ref{sec:prob} we have mentioned  various model dependent and independent bounds of NSIs and have discussed the
neutrino oscillation probability in presence of NSI. In section \ref{sec:sim}, we have discussed the procedure of numerical simulation
and in its different subsections related to superbeam, neutrino factory and monoenergetic neutrino beam we have discussed various experimental set-ups and results of our analysis on discovery of $CP$ violation. In section \ref{sec:con} we have concluded with
some remarks on the possibility of knowing the presence of NSIs in the context of superbeam and neutrino factory. For monoenergetic beam
we have discussed the suitable nuclei for discovering $CP$ violation and related technical constraints. 
\section{NSI and its effect on neutrino oscillation probabilities}
\label{sec:prob}
We consider the non-standard interactions of neutrinos which could be outcome of effective theory at low energy after integrating out the heavy mediator fields at the energy scale of neutrino oscillation experiments. Apart from  Standard Model (SM) Lagrangian density we consider the following non-standard fermion-neutrino interaction in matter  defined by the Lagrangian:

\bea
\label{eq:Lang}
\mathcal{L}_{NSI}^{M}=-2\sqrt{2}G_F\ve_{\a \b}^{f P}[\bar{f}\g_{\m}Pf][\bar{\n}_{\b}\g^{\m}L\n_{\a}]  
\eea
where $P \in (L,R)$, $L=\frac{(1-\g^5)}{2}$, $R=\frac{(1+\g^5)}{2}, $ $f=e, u, d$ and $\ve_{\a \b}^{fP}$ are termed as non-standard interactions (NSIs) 
parameters signifying the  deviation from SM interactions.  These are non-renormalizable as well as not gauge invariant and are dimension-6 operators after heavy fields are integrated out \cite{Ohlsson}. Although at low energy NSI look like this but at high energy scale
where actually such interactions originate there they have different form. These NSI parameters can be reduced to the effective parameters and can be written as:
\be
\label{eq:nsieffec}
\ve_{\a \b}=\sum_{f,P} \ve_{\a \b}^{fP}\frac{n_f}{n_e} 
\ee
where $n_f$  and $n_e$ are the fermion  and   the electron number density respectively in matter.
As these NSIs modify the interactions with matter from the Standard Model interactions the effective mass matrix for the neutrinos
are changed and as such there will be change in the oscillation probability of different flavor of neutrinos. Although NSIs could be present at the source of neutrinos, during the propagation of neutrinos and also during detection of neutrinos \cite{ra} but as those effects are  expected to be smaller at the source and
detector due to their stringent constraints \cite{nsi1,Ohlsson}, we consider the NSI effect during the propagation of neutrinos only. 

Model  dependent  \cite{nsi01p,nsi02p,nsi03p,nsi04p,nsi05p,nsi06p,nsi09p,nsi010p,nsi011p, indbound1,depbound2, 1loop, sknsi,Ohlsson} and independent \cite{nsi1,68,Es} bounds are obtained for these matter NSI parameters and are shown in the following table. In obtaining model dependent bounds on matter NSI the experiments with neutrinos and charged leptons - LSND, CHARM, CHARM-II, NuTeV and also LEP-II have been considered. Bounds coming from loop effect have been used for model dependent bounds. However, model independent bounds are less stringent and could be larger than the model dependent bounds 
by several orders and have been discussed in \cite{nsi1,Ohlsson}.  
\begin{table}[ht]
\centering 
\begin{tabular}{|c |c |c|} 

\hline 

NSI &  Model dependent & Model independent \\
& bound on NSI [Reference \cite{Ohlsson}]& Bound on NSI [Reference \cite{nsi1}]\\
\hline 
$\varepsilon_{ee}$ & $> -0.9; < 0.75$   & $ < 4.2$ \\
\hline
$|\varepsilon_{e \mu}|$ & $ \lsim 3.8 \times 10^{-4}$  & $ < 0.33$ \\
\hline
$|\varepsilon_{e \tau}|$ & $\lsim 0.25$  &  $< 3.0$ \\
\hline
$\varepsilon_{\mu \mu}$ & $  > -0.05 ;  <  0.08$  & $ < 0.068$ \\
\hline
$|\varepsilon_{\mu \tau}|$ & $ \lsim 0.25$  & $ < 0.33$ \\
\hline
$\varepsilon_{\tau \tau}$ & $ \lsim 0.4$  & $ < 21 $ \\
\hline
\end{tabular}
\caption{Strength of Non standard interaction terms used for our Analysis.} 
\label{table:bound} 
\end{table}
Considering recent results from experiments in IceCube-79 and DeepCore  more  stringent bound on $\varepsilon_{\mu \mu}$, 
$|\varepsilon_{\mu \tau}|$ and $\varepsilon_{\tau \tau}$ have been obtained in \cite{Es}.  
In section IV, we shall consider both  model dependent and independent allowed range of  values of different NSIs as shown in the table above for earth like matter  while showing discovery reach for $CP$ violation in presence of NSI. 

\subsection{\bf $\nu_e \rightarrow \nu_\mu$ oscillation probabilities with NSI}
The flavor eigenstates $\nu_\alpha$ is related to  mass eigenstates of neutrinos $\nu_i$ as
\be
\nu_\alpha =\sum_{i}  U_{\alpha i} \nu_i 
\; ;\;\quad 
\qquad i=1, 2, 3,
\ee
in vacuum where $U$ is  $3 \times 3$ unitary matrix  parametrized by
three mixing angles $\theta_{12}$, $\theta_{23}$ and $\theta_{13}$ and by one 
$CP$ violating phase $\delta$ for Dirac neutrino as given in \cite{bilen,part}. However, for Majorana neutrino apart from $\delta$ there are two more $CP$ violating phases 
which do not play role in neutrino oscillation \cite{bilen}. 
 
In the flavor basis the total Hamiltonian consisting both standard ($H_{SM} $) and non-standard interactions ($H_{NSI}$) of neutrinos interacting with matter during propagation can be written 
 as:
\begin{eqnarray}
\label{eq:hamil}
H &=& H_{SM} + H_{NSI} 
\end{eqnarray}
where
\begin{eqnarray}
\label{eq:H}
H_{SM} = \frac{\Delta m^2_{31}}{2E}\left[U\begin{pmatrix}0 & 0 & 0 \cr 0 & \alpha & 0 \cr
 0
 & 0 & 1\end{pmatrix} U^{\dag}+\begin{pmatrix}A & 0 & 0 \cr 0 & 0 & 0 \cr
 0
 & 0 & 0\end{pmatrix} \right], \nonumber \\
 \end{eqnarray}
 \begin{eqnarray}
 \label{eq:H1}
&&H_{\text{NSI}} =
\frac{\Delta m^2_{31}}{2E}\;A \begin{pmatrix}\varepsilon_{e e} & \varepsilon_{e \mu} & \varepsilon_{e \tau} \cr
\varepsilon_{e \mu}^* & \varepsilon_{\mu \mu} & \varepsilon_{\mu \tau} \cr
\varepsilon_{e \tau}^* & \varepsilon_{\mu \tau}^* & \varepsilon_{\tau \tau}\end{pmatrix} 
\end{eqnarray}
 In equations (\ref{eq:H}) and (\ref{eq:H1}) 
\bea
\label{eq:matternsi}
A = \frac{2E\sqrt{2}G_{F}n_{e}}{\Delta m_{31}^2} ; \; 
\alpha = \frac{\Delta m^2_{21}}{\Delta m^2_{31}} ; \;
\Delta m^2_{i j} = m^2_i - m^2_j \nonumber \\
\eea
where $m_i$ is the mass of the $i$-th neutrino, $A$ corresponds  to the interaction of neutrinos with
matter in SM and  $G_{F}$ is the Fermi constant. $\varepsilon_{ee}$, $\varepsilon_{e\mu}$ , $\varepsilon_{e\tau}$, $\varepsilon_{\mu\mu}$, $\varepsilon_{\mu\tau}$
 and $\varepsilon_{\tau\tau}$  correspond to the non-standard interactions (NSIs) of neutrinos with  matter. In equation (\ref{eq:H1}), ($\; ^{*} \;$)  
 denotes complex conjugation. The NSIs - $\ve_{e\mu}$, $\ve_{e\tau}$ and $\ve_{\mu\tau}$ could be complex. Later on, in the expressions of probability of oscillation we have
 expressed these NSIs as $\ve_{ij}=|\ve_{ij}|e^{i \phi_{ij}}$.
In   our numerical analysis we have considered the NSIs - $\ve_{e\mu}$, $\ve_{e\tau}$ and $\ve_{\mu\tau}$ as both real as well as complex.

For longer baselines following the perturbation method adopted in references \cite{m1,m2} for the oscillation probability  $P_{\nu_{e}\rightarrow \nu_{\mu}}$  upto order $\a^{2}$ (considering $\sin\theta_{13} \sim \sqrt{\a}$ as follows from recent reactor experiments)   and the  matter effect parameter  $A$ in the
leading order of perturbation (which happens for longer baselines) and NSI parameters $\ve_{\a\b} $ of the order of $\a$  one obtains \cite{m3}
\bea
\label{eq:prob}
P_{\nu_{e}\rightarrow \nu_\mu} &=& P^{SM}_{\nu_{e}\rightarrow \nu_\mu} + P^{NSI}_{\nu_{e}\rightarrow \nu_\mu}   
\eea
where
\begin{widetext}
\bea
\label{eq:prob1}
&&P^{SM}_{\nu_{e}\rightarrow \nu_\mu} = 4\sin \frac{(A-1)\Delta m^2_{31}L}{4E}\frac{s^2_{13}s^2_{23}}{(A-1)^4}\bigg(((A-1)^2-(1+A)^2s^2_{13})\sin\frac{(A-1)\Delta m^2_{31}L}{4E}\nonumber \\
&+&A(A-1)\frac{\Delta m^2_{31}L}{E}s^2_{13}\cos\frac{(A-1)\Delta m^2_{31}L}{4E}\bigg)
+ \frac{\a^2c^2_{23}}{A^2}\sin^2 2\theta_{12}\sin^2\frac{\Delta m^2_{31}AL}{4E}\nonumber \\
&+&\frac{\a s^2_{12}s^2_{13}s^2_{23}}{(A-1)^3}\bigg(\frac{(A-1)\Delta m^2_{31}L}{E}\sin \frac{(A-1)\Delta m^2_{31}L}{2E}-8A\sin^2 \frac{(A-1)\Delta m^2_{31}L}{4E}\bigg)\nonumber \\
&+& \frac{\a s_{13}s_{2 \times 12}s_{2 \times 23}}{A(A-1)}\bigg(2\cos\bigg(\delta - \frac{\Delta m^2_{31}L}{4E}\bigg)\sin \frac{(A-1)\Delta m^2_{31}L}{4E}\sin\frac{A\Delta m^2_{31}L}{4E}\bigg) 
\eea
\end{widetext}
\begin{widetext}
\bea
\label{eq:pro2}
P^{NSI}_{\nu_{e}\rightarrow \nu_\mu} &&= \frac{4|a_2|s_{2\times 23}s_{13}}{A(A-1)}\sin \frac{A\Delta m^2_{31}L}{4E}\sin\frac{(A-1)\Delta m^2_{31}L}{4E}\cos \big(\delta - \frac{\Delta m^2_{31}L}{4E}+\phi_{a_2}\big)\nonumber \\
&+& \frac{4|a_3|s^2_{23}}{(A-1)^2}\sin^2 \frac{(A-1)\Delta m^2_{31}L}{4E}(|a_3|+2\cos(\delta + \phi_{a_3})s_{13})\nonumber \\
&+&\frac{s^2_{13}s^2_{23}(|a_5|-|a_1|)}{(A-1)^3E}\bigg(8E\sin^2 \frac{(A-1)\Delta m^2_{31}L}{4E}-(A-1)\Delta m^2_{31}L\sin \frac{(A-1)\Delta m^2_{31}L}{2E}\bigg)\nonumber \\
&+&\frac{4 |a_2|c_{23}}{(A-1)A^2}\sin \frac{A\Delta m^2_{31}L}{4E}\bigg((A-1)c_{23}\sin \frac{A\Delta m^2_{31}L}{4E}(|a_2|+\a\cos \phi_{a_2}\sin 2\theta_{12})\bigg)\nonumber \\
&-&\frac{4|a_2| |a_3|\sin 2\theta_{23}}{A(A-1)}\cos\bigg[\frac{\Delta m^2_{31}L}{4E}-\phi _{a_2}+\phi_{a_3}\bigg]\sin \frac{(1-A)\Delta m^2_{31}L}{4E}\sin \frac{A\Delta m^2_{31}L}{4E}\nonumber \\
&+& \frac{4|a_3|s_{23}}{(A-1)^2A}\sin \frac{(A-1)\Delta m^2_{31}L}{4E}(A-1)\a c_{23}\cos\bigg[\frac{\Delta m^2_{31}L}{4E}-\phi _{a_3}\bigg]\sin \frac{A\Delta m^2_{31}L}{4E}\sin 2\theta_{12}\nonumber \\
&+&\frac{|a_4|s^2_{13}\sin 2 \theta_{23}}{(A-1)^2A}\sin \frac{(A-1)\Delta m^2_{31}L}{4E}\bigg(-4A\cos\frac{A\Delta m^2_{31}L}{4E} \cos \phi_{a_4}\sin \frac{\Delta m^2_{31}L}{4E}\nonumber\\
&+&4\sin \frac{A\Delta m^2_{31}L}{4E}\bigg(\cos\frac{\Delta m^2_{31}L}{4E}\cos \phi_{a_4}+(A-1)\sin\frac{\Delta m^2_{31}L}{4E}\sin\phi_{a_4}\bigg)\bigg) 
\eea
\end{widetext}
where
\bea
&&a_1 = A \ve_{e e} \nonumber \\
&&|a_2|e^{i \phi_{a_2}} = A \bigg(e^{i\phi_{e\mu}}|\ve_{e\mu}|c_{23}-e^{i\phi_{e\tau}}|\ve_{e\tau}|s_{23}\bigg) \nonumber \\ 
&&|a_3|e^{i \phi_{a_3}} = A \bigg(e^{i\phi_{e\tau}}|\ve_{e\tau}|c_{23}+e^{i\phi_{e\mu}}|\ve_{e\mu}|s_{23}\bigg) \nonumber \\
&&|a_4|e^{i \phi_{a_4}} = A \bigg(|\ve_{\mu \tau}|e^{i\phi_{\mu \tau}} - 2 |\ve_{\mu \tau}|s^2_{23} + (\ve_{\mu \mu}-\ve_{\tau \tau})c_{23}s_{23}\bigg) \nonumber \\
&&a_5 = A \bigg(  \ve_{\tau \tau} c^2_{23} +  \ve_{\mu \mu} s^2_{23} +  |\ve_{\mu \tau}| \cos \phi_{\mu \tau}s_{2 \times 23}\bigg) \nonumber \\
\eea

and
\bea
\phi_{a_2}&=&\tan^{-1}\left[\frac{\vert\ve_{e \mu}\vert c_{23}  \sin \phi_{e\mu} -\vert\ve_{e \tau}\vert s_{23}  \sin \phi_{e \tau}}{\vert\ve_{e \mu}\vert c_{23}  \cos \phi_{e\mu}]-\vert\ve_{e \tau}\vert \cos \phi_{e \tau}] s_{23} }\right]\nonumber \\
\phi_{a_3}&=&\tan^{-1}\left[\frac{\vert\ve_{e\mu}\vert s_{23}  \sin \phi_{e \mu}+\vert\ve_{e \tau}\vert c_{23} \sin \phi_{e \tau}}{\vert\ve_{e \tau}\vert c_{23}  \cos \phi_{e \tau} +\vert\ve_{e\mu}\vert \cos \phi_{e \mu} s_{23} }\right] \nonumber \\
\phi_{a_4} &=& \tan^{-1}\left( \frac{|\varepsilon_{\mu \tau}|\sin(\phi_{\mu \tau})}{|\varepsilon_{\mu \tau}|c_{2\times 23}\cos(\phi_{\mu \tau})+(\varepsilon_{\mu \mu} - \varepsilon_{\tau \tau})c_{23} s_{23}}\right) \nonumber \\
\eea
where $s_{ij}=\sin \theta_{ij}$, $c_{ij}=\cos \theta_{ij}$, $s_{2 \times ij}=\sin 2\theta_{ij}$, $c_{2 \times ij}=\cos 2\theta_{ij}$. 
The oscillation channel $ \nu_e \rightarrow \nu_\mu $ is important for neutrino factory. For superbeam, $\nu_\mu \rightarrow \nu_e $ 
 oscillation channel is important which can be obtained by the following transformation 
 \begin{equation}
P_{{\alpha}{\beta}} (A, \delta, \phi_{ij})= P_{ \beta \alpha}(A, -\delta, -\phi_{ij}).
\label{nufacttosb}
\end{equation}
The oscillation  probabilities for antineutrinos can be obtained from the oscillation probabilities given for neutrinos above  
by using the following relation:
\begin{equation}
P_{\bar{\alpha}\bar{\beta}}(\delta  ,\; {A}, \; \phi_{ij} )= P_{\alpha \beta}(-\delta  ,\; {-A}, \; -\phi_{ij} ).
\label{probsd}
\end{equation}

To estimate the order of magnitude of $\delta $ dependent and $\delta $ independent  but matter dependent ( i.e., $A $ dependent) part in the above two oscillation probability,  
following results of reactor experiments  we shall consider $\sin\theta_{13} \sim \sqrt{\a}$. For only SM interactions, (i.e  $\ve_{\a\b} \rightarrow 0$) from  last term of Eq.\eqref{eq:prob1}  for oscillation probabilities one finds that  the 
$\delta$ dependence occurs at order of $ \a s_{13} \sim \a^{3/2}$ for both neutrino oscillation and  anti-neutrino oscillation probabilities. 

In presence of NSI of order $\a$, in the first two leading terms of Eq.\eqref{eq:pro2}, the  $\delta$ dependence occurs at order of $ a_2 s_{13} \sim a_3 s_{13} \sim \a^{3/2}$ through terms containing $a_2$ and $a_3$ (which are $\ve_{e\mu}$ and $\ve_{e\tau}$ dependent). 
Due to this (if we assume all NSI of same order $\sim \a $),  these two NSI ($\ve_{e\mu}$ and $\ve_{e\tau}$) in contrast to other NSI could have greater effect on oscillation probability in $\nu_e \rightarrow \nu_\mu$ or the reverse oscillation 
channel .

For shorter baselines the above oscillation probability expression becomes much simpler. Following the perturbation method \cite{ra} and by considering the standard model matter effect $A \sim \alpha$  for neutrino energy $E$ around 1 GeV  
and  $\sin\theta_{13} \sim \sqrt{\a}$ as obtained from reactor oscillation data,  the probability of oscillation $P_{\nu_e \rightarrow \nu_{\mu}}$  upto order $\alpha^{2}$ is given by

\begin{widetext}
 \begin{eqnarray}
  P_{\nu_e \rightarrow \nu_{\mu}} = P_{\nu_e \rightarrow \nu_{\mu}}(\alpha) + P_{\nu_e \rightarrow \nu_{\mu}}(\alpha^{3/2}) + P_{\nu_e \rightarrow \nu_{\mu}}(\alpha^2) 
\label{eq:p250}
\end{eqnarray}
\end{widetext}
where 
\begin{widetext}
 \begin{eqnarray}
 P_{\nu_e \rightarrow \nu_{\mu}}(\alpha)&=& \sin^2 2\theta_{13}\sin^2 \theta_{23}\sin^2\bigg(\frac{\Delta m^2_{31}L}{4E}\bigg) \nonumber \\
  P_{\nu_e \rightarrow \nu_{\mu}}(\alpha^{3/2}) &= &\alpha\bigg(\frac{\Delta m^2_{31}L}{2E}\bigg) \sin \theta_{13}\sin 2\theta_{23}\sin 2 \theta_{12}\sin \bigg(\frac{\Delta m^2_{31}L}{4E}\bigg)
 \cos \bigg(\delta - \frac{\Delta m^2_{31}L}{4E}\bigg) \nonumber \\
 P_{\nu_e \rightarrow \nu_{\mu}}(\alpha^2)&=&  \a^2 \cos^2\theta_{23}\sin^2 2\theta_{12}\bigg(\frac{\Delta m^2_{31}L }{4E}\bigg)^2-2\a \sin^2 \theta_{13}\sin^2 \theta_{12}\sin^2 \theta_{23}\frac{\Delta m^2_{31}L}{2E}
 \sin \frac{\Delta m ^2_{31}L}{2E}  \nonumber \\ &+& 8A\sin^2 \theta_{13}\sin^2 \theta_{23}\bigg(\sin^2 \frac{\Delta m ^2_{31}L}{4E}-\frac{\Delta m ^2_{31}L}{8E} \sin \frac{\Delta m ^2_{31}L}{2E}\bigg)
\label{eq:p250}
\end{eqnarray}
\end{widetext}
 One may note that this expression of oscillation probability is little bit different  from that presented by earlier authors \cite{akh} because they have considered the perturbative approach for relatively longer baseline and  small $\sin\theta_{13} \sim \alpha$. 
 For shorter baseline as presented here, the $\delta$ dependence appears to be more and instead of being at the order of $\alpha^2$ it appears at the order $\alpha^{3/2}$ as $ \a \sin\theta_{13} \sim {\a}^{3/2}$ although the expression is same. However at order $\alpha^2$ some extra terms appear in comparison to that presented by earlier authors although their contribution is relatively smaller being at the order of $\alpha^2$.
The matter effect  is occurring at order $\a^2$ through term containing parameter $A$. 
\section{Numerical simulation}
\label{sec:sim}
There is detailed global analysis of three flavor neutrino oscillation data \cite{Fogli:2012ua} presented at the \emph{Neutrino 2012} conference. In their analysis the correlation between various oscillation parameters has been taken into account.
Very recently there is another such global analysis \cite{Gonzalez-Garcia:2014bfa} presented at the \emph{Neutrino 2014} conference which
has been considered in our analysis and best fit values of neutrino mixing parameters and their respective errors at $3\sigma$ confidence level  are shown
in  table
\ref{table:mix}.
For earth matter
density the PREM profile \cite{prem} has been considered. Also we have considered an error of $2\%$ on matter density profile.

\begin{table}[ht]
\centering 
\begin{tabular}{|c |c |c |} 
\hline 
Oscillation Parameters & Central Values & Error ($3\sigma$)  \\
\hline
$\frac{\Delta m^2_{21}}{10^{-5}} \; \textrm{eV}^2$ & $7.45 $ & 7.02 - 8.09 \\
\hline
\multirow{2}{*}{$\frac{\Delta m^2_{31}}{10^{-3}} \; \textrm{eV}^2$} &\multirow{2}{*}{ $2.457$} & 2.325 - 2.599 (NH)  \\
& & (-2.59)- (-2.307) (IH)\\
\hline
$\sin^2 \theta_{12}$ & 0.304 & 0.270 - 0.344  \\
\hline
$ \sin^2 \theta_{23}$ & 0.452 & 0.385 - 0.644  \\
\hline
$\sin^2 \theta_{13}$ & 0.0218  & 0.0188 - 0.0251  \\
\hline
\end{tabular}
\caption{Central values of the oscillation parameters with errors.} 
\label{table:mix} 
\end{table}

The numerical simulation has been done by using GLoBES \cite{globes1,globes11}. GLoBES uses poissonian $\chi^2$ for the 
    oscillation parameters ${\bf \lambda}$ and for nuisance parameters $\xi_i$.  To implement  the systematic errors the
    `pull method' \cite{pull} has been used by GLoBES and $\chi^2_{pull}$ is given as  \cite{globes11}
\begin{align}
\chi^2_{pull}({\bf \lambda}) &:= \min_{\{\xi_i\}}\bigg(\chi^2({\bf \lambda},\xi_1,\cdots,\xi_k)+\sum_{j=1}^k\frac{\xi^2_j}{\sigma^2_{\xi_j}}\bigg) 
\end{align}
where $\chi^2({\bf \lambda},\xi_1,\cdots,\xi_n)$ is the usual Poissonian $\chi^2$ depending on the neutrino oscillation parameters
$\lambda$ and the nuisance parameter $\xi_i$ and it has been summed over different energy binning on the  theoretical and observed event rates. The oscillation parameters ${\bf \lambda}$ is 
\begin{align}
{\bf \lambda} &= (\theta_{12},\theta_{13},\theta_{23},\Delta m^2_{21},\Delta m^2_{31},\rho,\delta_{CP},\varepsilon_{\alpha \beta},\phi_{\alpha \beta})\; ;
\end{align}
The nuisance parameters $\xi_i$ are the signal and background normalization and the calibration errors. In order to implement the error $\sigma_{\xi_i}$, the $\chi^2$ (in which all
event rates of all bins are included) is minimized over the different nuisance parameters $\xi_i$ independently.

      We have also marginalized over ${\bf \lambda^\prime}$ where 
\begin{align}
{\bf \lambda^\prime} &= (\theta_{12},\theta_{13},\theta_{23},\Delta m^2_{21},\Delta m^2_{31},\rho)\; .
\end{align}
The final projected $\chi_F^2$ is given as
\begin{align}
\chi_F^2 &= \min_{\lambda^\prime} \left( \chi^2_{pull}(\lambda) + priors (\lambda^\prime)\right)
\end{align}
where
\begin{align}
priors(\lambda^\prime) &= \bigg(\frac{\rho - \rho^0}{\sigma_{\rho}}\bigg)^2 + \sum_{i\neq j} \bigg(\frac{\sin^2\theta_{ij} - \sin^2\theta_{ij}^0}{\sigma_{\sin^2\theta_{ij}}}\bigg)^2 + \bigg(\frac{\Delta m^2_{21} -  (\Delta m^{2}_{21})^0}{\sigma_{\Delta m^2_{21}}}\bigg)^2 + \bigg(\frac{|\Delta m^2_{31}| - | (\Delta m^{2}_{31})^0|}{\sigma_{|\Delta m^2_{31}|}}\bigg)^2
\end{align}
and $\rho^0,\; \sin^2\theta^0_{12},\; \sin^2\theta^0_{13},\;\sin^2\theta^0_{23},\; (\Delta m^2_{21})^0 \text{and} | (\Delta m_{31}^2)^0|)$ are the 
central values of the corresponding parameters. We have marginalized over two different hierarchies of neutrino masses also.

For taking into account NSI  in GLoBES we have followed the method described in \cite{globes11} and modified the source file "probability.c" in GLoBES appropriately
by inserting the NSI's in the subroutine where the hamiltonian for matter interaction is defined. Then we have included the new probability program as instructed by the manual of GLoBES. We have not used  the approximate analytical expression of oscillation probability for NSI of order $\a$ in our numerical 
analysis.

At present we do not know the hierarchy of neutrino masses and the value of  $\delta$. In the test values we have considered both
possibilities of the hierarchy. However, we shall show the results in Figures only for true normal hierarchy. Before we get the knowledge of specific value of $\delta$  it could be possible to know
whether nature admits $CP$ violation or not which could be possible for a wide range of values of $\delta$. So what we are going to analyse
is not the measurement of $\delta$ values but finding out all possible $\delta$ (true) values which could be distinguishable from
the  $CP$ conserving $\delta $ values at certain confidence level through oscillation experiments. 
For that it is useful to calculate the $CP$ fraction due to $\delta $ denoted as $F(\delta )$ which  may be defined as the fraction of the total allowed 
range (0 to 2 $\pi$) for the  $\delta$(true) over which $CP$ violation can be discovered \cite{nsi3,Ballett:2012rz}.  While calculating the fraction due to
$\delta$ for the discovery of $CP$ violation, we consider the exclusion of all parameter sets with $\delta \in \{0,\pi \}$ which means
$\delta$(test) is fixed in its $CP$ conserving values $\{0,\pi\}$. For no NSIs, the degrees of freedom in $\chi^2$ analysis is one due to
due to $\delta$.
 We have calculated the minimum of those $\chi^2$ for each $\delta$(true) and have calculated the fraction of the $\delta$(true) over its entire range which satisfies $\Delta \chi^2 \geq 9 ( 25) \; \text{at} \;  3 (5) \sigma$ confidence level.

In studying the NSI effect (one at a time) on $CP$ violation we have fixed absolute values of NSI(test)  to their true values
which implies that the absolute values of NSI are assumed to be known from some other experiments. 
 In 
Figures we have shown the effect of real NSI's  on
the $CP$ fraction discovery. In this case, the degrees of freedom in $\chi^2$ analysis is one due to $\delta$.
The off-diagonal
NSI in NSI matrix could be complex and the corresponding complex phases could be new sources of $CP$ violation. If the $CP$ violation
is only due to NSI phase, then the $CP$ fraction due to NSI phases ${\phi_{ ij}}$  (denoted as $F({\phi_{ ij}})$) is calculated in the same way as described above for $\delta$ by replacing $\delta$ by the corresponding $\phi_{ ij}$.   Here the degrees of freedom is one 
due to $\phi_{ij}$.   In 
Figures we have shown the effect of absolute values of non-diagonal NSI's (one at a time) on
$F({\phi_{ ij}})$. In order to calculate the the total $CP$ violation  discovery due to two phases $\phi_{ ij}$ and $\delta$  we have
 set $\phi_{ ij}$(test) and $\delta$(test) to their respective $CP$ conserving values at $(\delta,\phi_{ij})\in \{(0,0),(0,\pi),(\pi,0),(\pi,\pi)\}$  and also  both of the mass hierarchy in the test values have been taken into account.  Here, the degrees of freedom is two due to $\delta $ and  one of the NSI phases $\phi_{ij}$. Then we have calculated the above $\chi^2$ minimum for each true values of 
$\phi_{ ij}$ and $\delta$ and have shown the allowed region for discovery in $\phi_{ ij}$-$\delta$ space for $\Delta \chi^2 \geq 11.83
$ at $3 \sigma$ confidence level.
In all Figures true hierarchy has been considered to be normal.

So by numerical simulation
we are studying the different possibilities of finding $CP$ violation in different experimental set-ups due to $\delta$ or NSI phase $\phi_{ ij}$ or total
$CP$ violation due to $\delta $ and NSI phase $\phi_{ ij}$ for  certain absolute value of NSI ($\ve_{ij}$ (true)) assumed to be present in nature.   If we assume that any allowed values of $\delta $ has equal probability to be true $\delta$ value of nature then 
lower (higher) $CP$
fraction which we present in the result sections will indicate lower (higher) probability of finding $CP$ violation in particular 
experimental-set-up discussed below. One may note that NSI effect and the $\delta $ effect in the hamiltonian $H$  has different neutrino energy dependence (see equation Eq.\eqref{eq:H} for $\delta$ present in $U$ and Eq.\eqref{eq:H1} for NSI). Also 
in the shorter (longer) baseline there should be lesser (larger) NSI effect.  If the experimental data is available over certain range of neutrino energy from the shorter and longer  baseline experiments then only the multi-parameter fit
will help in disentangling  effects due to different unknown parameters and hence improving bounds on NSI parameters or
discovering it.

In the following we present three subsections based on three different sources  of neutrino flux :  Superbeam in \ref{sec:sb}, 
 Neutrino Factory in \ref{sec:nu} and  monoenergetic neutrino beam in \ref{sec:mono}. Particularly for Superbeam and 
 Neutrino factory, we have studied the NSI effect 
 in $CP$ violation discovery in the leptonic sector in  neutrino oscillation experiments. Monoenergetic neutrino beam requires shorter baseline due to technical reasons as discussed later. Due to that the NSI effect in the monoenergetic neutrino beam studies is insignificant and we will be discussing only $CP$ violation discovery reach in absence of NSI. Two different nuclei have been considered for
 monoenergetic beam and to study  $CP$ violation.

\subsection{\bf Superbeam}
\label{sec:sb}
We discuss below the experimental set-up and systematic errors for superbeam.  We have presented our results for discovery 
reach of $CP$ violation 
due to $\delta $ and other off-diagonal NSI phases $\phi_{ij}$ for different experimental set-up with 
different baselines. We have shown also the effect of the absolute values of different $\varepsilon_{\a\b}$ in $CP$ fraction.

\subsubsection{\bf Experimental set-ups and systemetic errors}
We consider three experimental set-ups:
 (a) A Superbeam set-up originating in CERN and reaching a 500 kt Water Cherenkov detector 
 \cite{Agostino:2012fd} placed at a distance of 130 km at Fr\'ejus (France), (b) A Superbeam set-up originating in Tokai and reaching a 500 kt Water Cherenkov detector  \cite{Huber:2002mx,Itow,Ishitsuka:2005qi} placed at a distance of 295 km at Kamiokande (Japan) and (c) A Superbeam set-up which originates in CERN and reaches a 100 kt Liquid Argon detector placed at a distance of 2300 km at Pyh\"asalmi (Finland).
The observable channels that we have considered are 
$\nu_{\mu}\rightarrow \nu_e$, $\bar{\nu}_{\mu}\rightarrow \bar{\nu}_e$, $\nu_{\mu}\rightarrow \nu_{\mu}$ and $\bar{\nu}_{\mu}\rightarrow \bar{\nu}_{\mu} $. 

The flux considered for set-up (a) has  mean energy $\sim 0.3$ GeV, which are computed for 3.5 GeV protons and $10^{23}$ protons on target per year. For our analysis the beam power has been considered of about 4 MW per year and the time period has been taken to be 2 years for neutrinos and 4 years for anti-neutrinos.
We consider the same flux as in \cite{others5,splflux}.
In the case of set-up(a)  the efficiencies for the signal and background are 
included in the migration matrices based on \cite{Agostino:2012fd} except for the channels $\nu_\mu$ disappearance, $\bar{\nu}_\mu$ 
disappearance and $\nu_\mu$ (NC) which are $64\%$, $81\%$ and  $11.7\%$ efficiencies respectively. We have considered systematic uncertainties of $2\%$ on signal and background channels.

The experimental set-up (b) has been considered as given in GLoBES \cite{hkflux} and
the detector  specification is 
mentioned in table \ref{table:hk}:
\begin{table}[h!]
\centering
\begin{tabular}{|c|c|}
\hline 
Target Power & 4 Mega-Watt \\
	 \hline
	 Fiducial Mass & 500 kt \\
	   \hline
	   Data Taking & 4 yrs $\nu$ and 4 yrs $\overline{\nu}$ \\
	     \hline
	     Baseline & 295 km \\
	      \hline	       
	       Energy Resolution & 0.085 \\
		 \hline
		 \end{tabular}
\caption{Detector characteristics for Hyper-Kamiokande.} 

\label{table:hk} 

		 \end{table}

The flux considered for set-up (c) has  mean energy $\sim 5$ GeV, which are computed for 50 GeV protons and $3\times 10^{21}$ protons 
on target per year. For our analysis the beam power is about 0.8 MW per year and the time period has been taken to be 5 years each for neutrinos and anti-neutrinos.
We consider the same flux as in \cite{Coloma:2012ma}. The detector characteristics for set-up (c) is given in table \ref{table:charac} \cite{Adams:2013qkq}.
The correlation between the visible energy of background NC events and the neutrino energy is
implemented by migration matrices which has been  provided by L. Whitehead \cite{white}.

\begin{table}[ht]

\centering 

\begin{tabular}{|c |c |c |c|} 

\hline
\hline 
 Signal Studies& $\nu_e$ CC appearance Studies & $\nu_\mu$ CC Disappearance Studies  \\
\hline
Signal efficiency &  80\% & 85\%  \\
\hline
$\nu_\mu$ NC mis-identification rate (Background) & 1\%  & 0.5\% \\
\hline
$\nu_\mu$ CC mis-identification rate (Background) & 1\% & 0\% \\
\hline 
Signal Normalization error & 5\% & 10\% \\
\hline
Background Normalization Error & 15\% & 20\% \\
\hline
\multicolumn{3}{|c|}{Neutrino Energy Resolution} \\
\hline
\hline
$\nu_e$ CC energy resolution & \multicolumn{2}{|c|}{0.15$\sqrt{E(\textrm{GeV})}$} \\
\hline
$\nu_\mu$ CC energy resolution & \multicolumn{2}{|c|}{0.2$\sqrt{E(\textrm{GeV})}$} \\
\hline
$E_{\nu_\mu}$ scale uncertainty &  \multicolumn{2}{|c|}{2\%} \\
\hline
$E_{\nu_e}$ scale uncertainty &  \multicolumn{2}{|c|}{0.01\%} \\
\hline
\end{tabular}
\caption{Detector characteristics for set-up (c).} 

\label{table:charac} 

\end{table}

\begin{table}
\begin{tabular}{|c|c|c|c|}
\hline
\hline
 & {CERN-Fr\'ejus } & {T2HK } & {CERN-Pyh\"asalmi } \\
&(130 km) &(295 km) & (2300 km) \\
 \hline
SM & 4533 & 6098 & 421 \\
\hline
$\varepsilon_{ee}$ & 4536 & 6110 &  426 \\
\hline
$\varepsilon_{e\mu}$ &4634 & 6272 & 502  \\
\hline
$\varepsilon_{e\tau}$ & 4543 & 6143  & 470 \\
\hline
$\varepsilon_{\mu\tau}$ & 4529 & 6081 & 414 \\
\hline
$\varepsilon_{\mu\mu}$ & 4533 & 6099 & 425 \\
\hline
$\varepsilon_{\tau\tau}$ & 4530 & 6085  & 413 \\
\hline
\end{tabular}
\caption{The number of $\nu_{\mu}\rightarrow \nu_e$ events for CERN-Fr\'ejus (130 km), T2HK (295 km) and CERN-Pyh\"asalmi (2300 km)  for SM and for different NSI (one at a time each of which equals to $\a$).}
\label{table:eventssb} 
\end{table}
In table \ref{table:eventssb} for $\delta =0$ we have shown the expected number of events for three experimental set-ups, for no NSIs and also for different
real NSIs , each of which equal to $\a$ (one NSI at a time). We have considered the central values of various parameters
as shown in table \ref{table:mix} and have used PREM profile \cite{prem} for matter densities. 
One can see that there is significant effect of  $\ve_{e\mu}$  and $\ve_{e\tau}$ on the number of $\nu_{\mu}\rightarrow \nu_e$ events.  
Possibilities of significant effect of these NSIs can be seen in the expression of oscillation probabilities in Eq.\eqref{eq:pro2}.

\subsubsection{\bf Results}
\begin{figure}[h!]
\centering
\begin{tabular}{cc}
\includegraphics[width=0.5\textwidth]{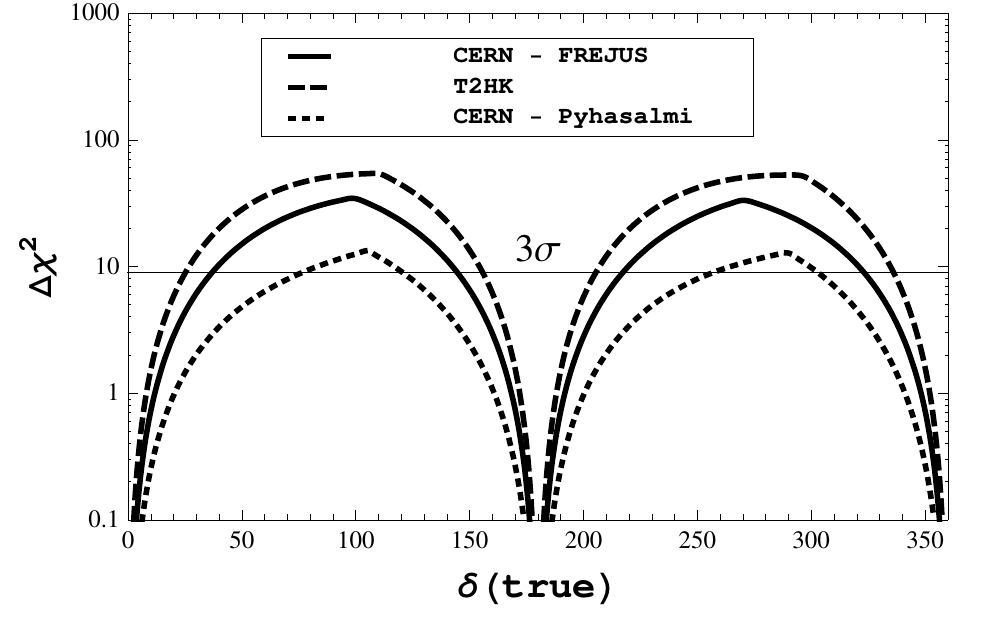}
\end{tabular}
\caption[] {{\small Discovery reach of $CP$ violation due to $\delta$ for three different experimental set-up :  CERN-Fr\'ejus, T2HK and CERN-Pyh\"asalmi set-up considering only SM interactions for normal hierarchy. }}
\label{fig:delta1}
\end{figure}

\begin{figure}[htb]
\centering
\begin{tabular}{cc}
\includegraphics[width=0.45\textwidth]{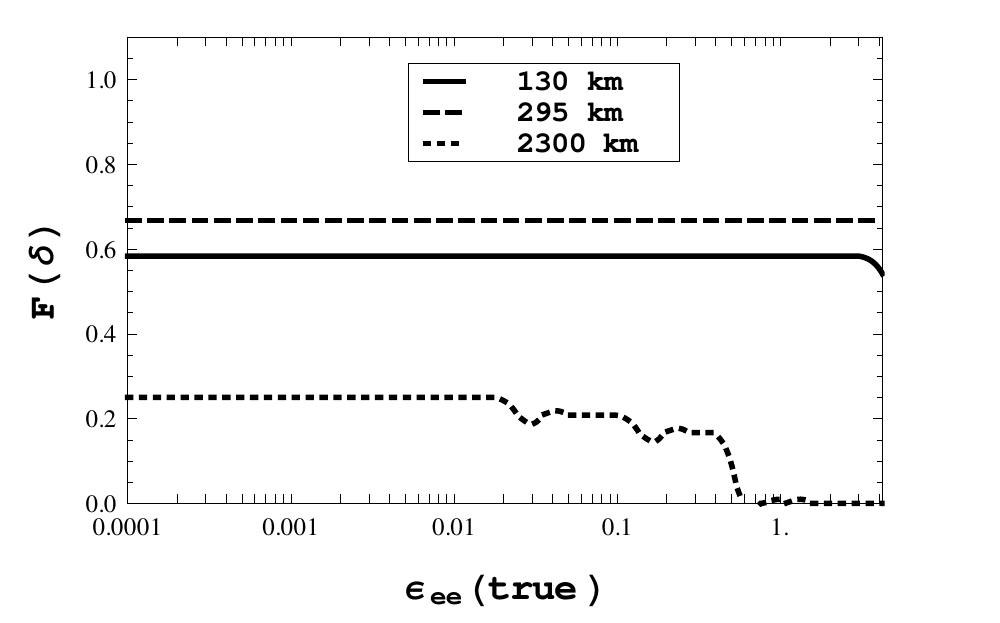}&
\includegraphics[width=0.45\textwidth]{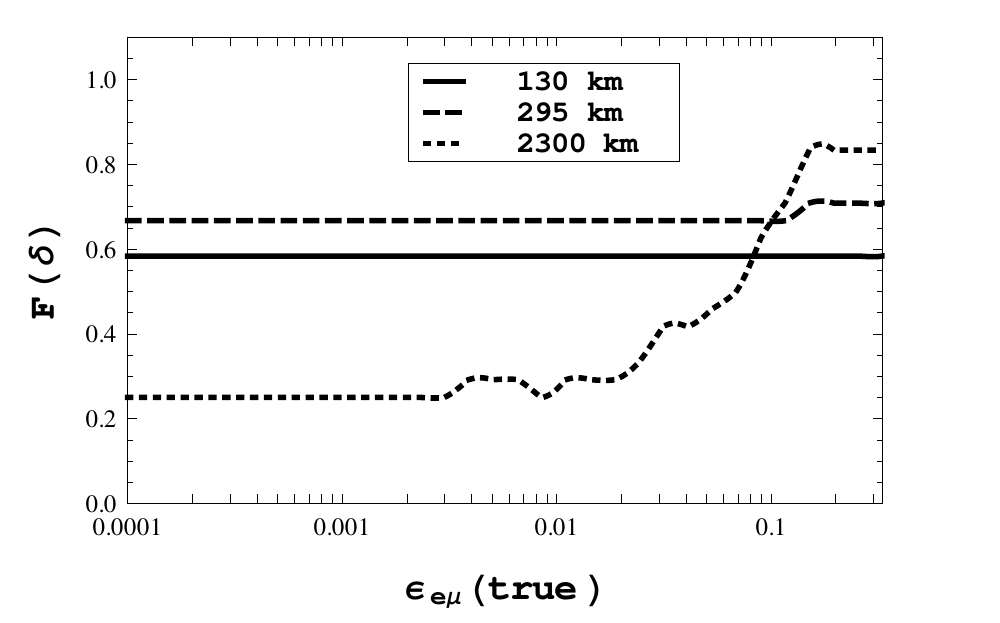}\\
\includegraphics[width=0.45\textwidth]{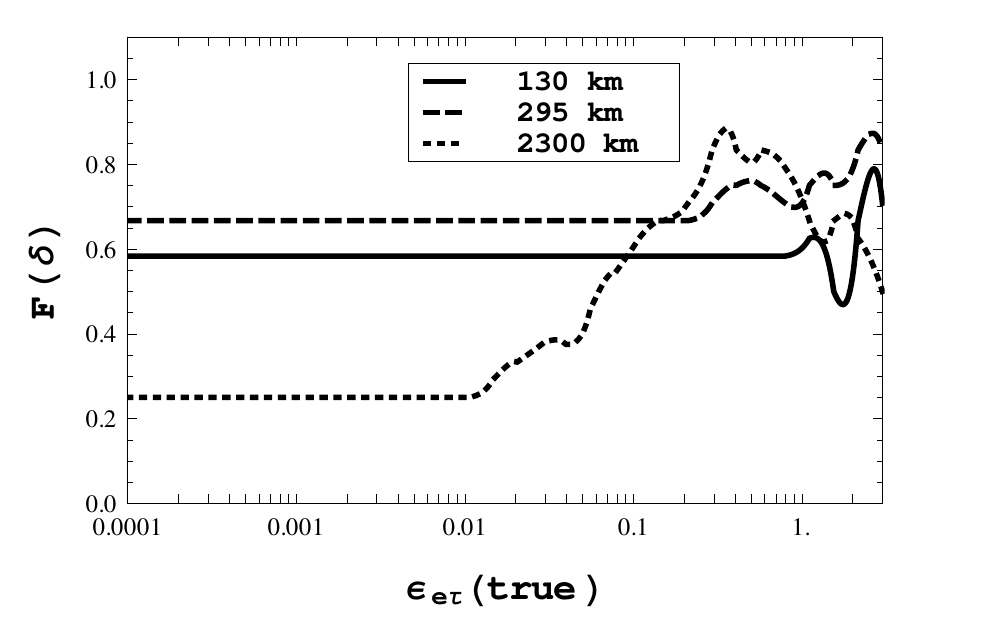}&
\includegraphics[width=0.45\textwidth]{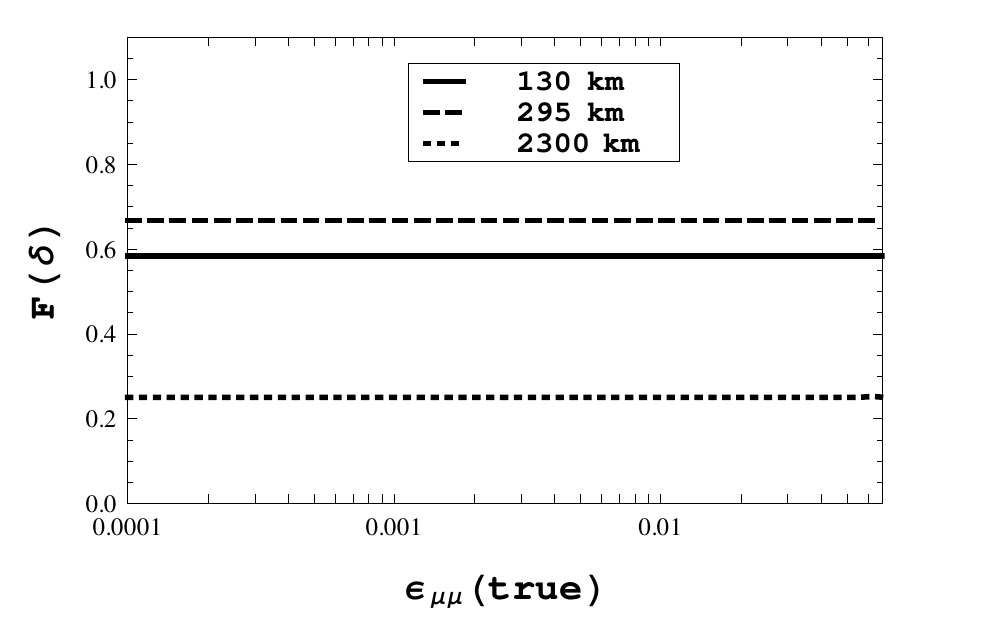}\\
\includegraphics[width=0.45\textwidth]{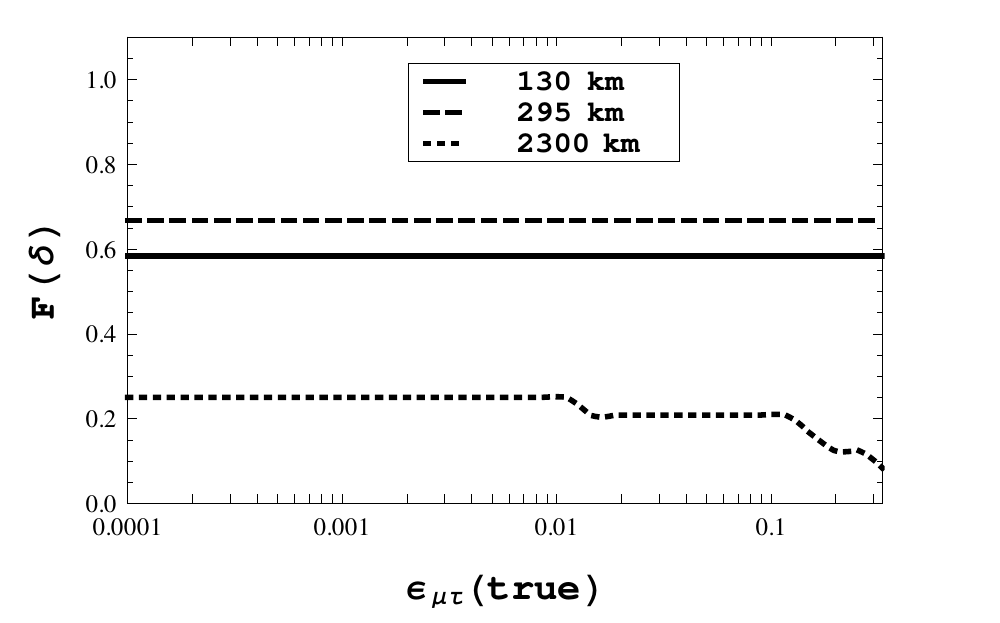}&
\includegraphics[width=0.45\textwidth]{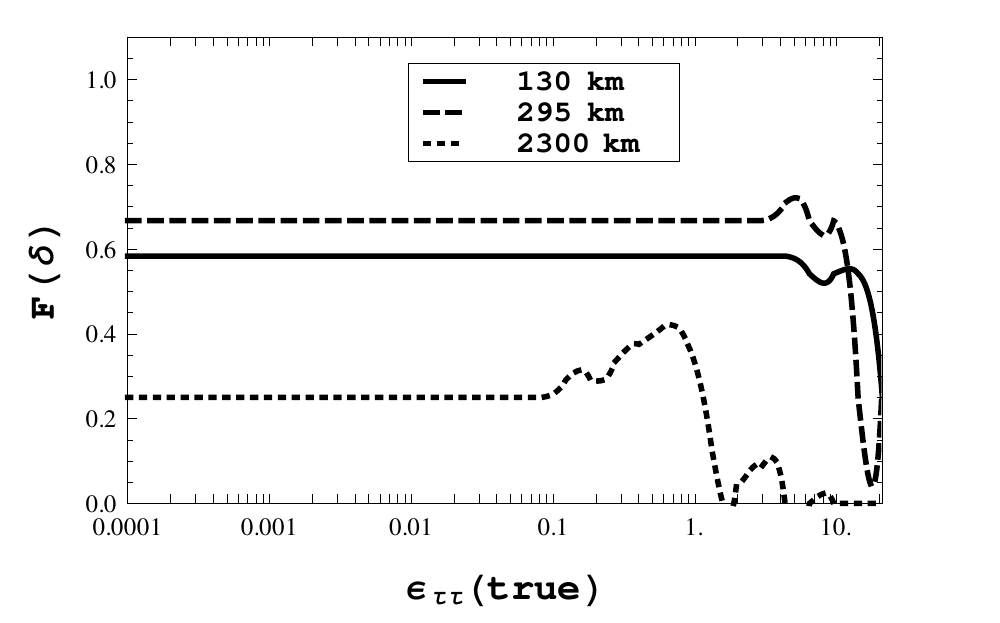} 
\end{tabular}
\caption[] {{\small $F(\delta)$ for three different experimental set-up:  CERN-Fr\'ejus, T2HK and CERN-Pyh\"asalmi set-up at $3\sigma$ considering real NSIs $\ve_{ee}$, $\ve_{e\mu}$ , $\ve_{e\tau}$, $\ve_{\mu\mu}$, $\ve_{\mu\tau}$ and $\ve_{\tau\tau}$. }}
\label{fig:delta2}
\end{figure}

\begin{figure*}[htb]
\centering
\begin{tabular}{cc}
\includegraphics[width=0.45\textwidth]{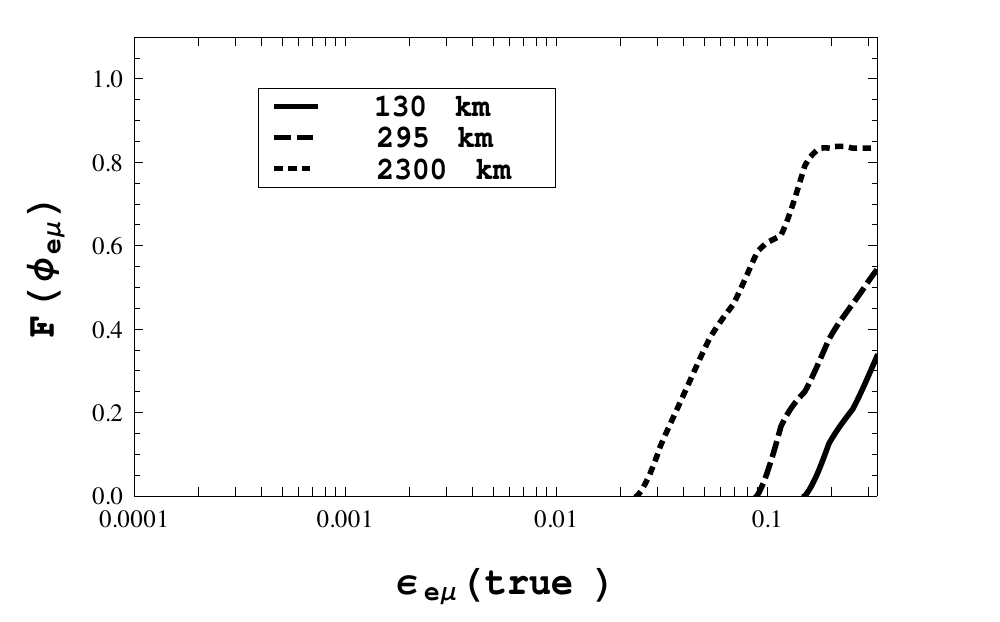}&
\includegraphics[width=0.45\textwidth]{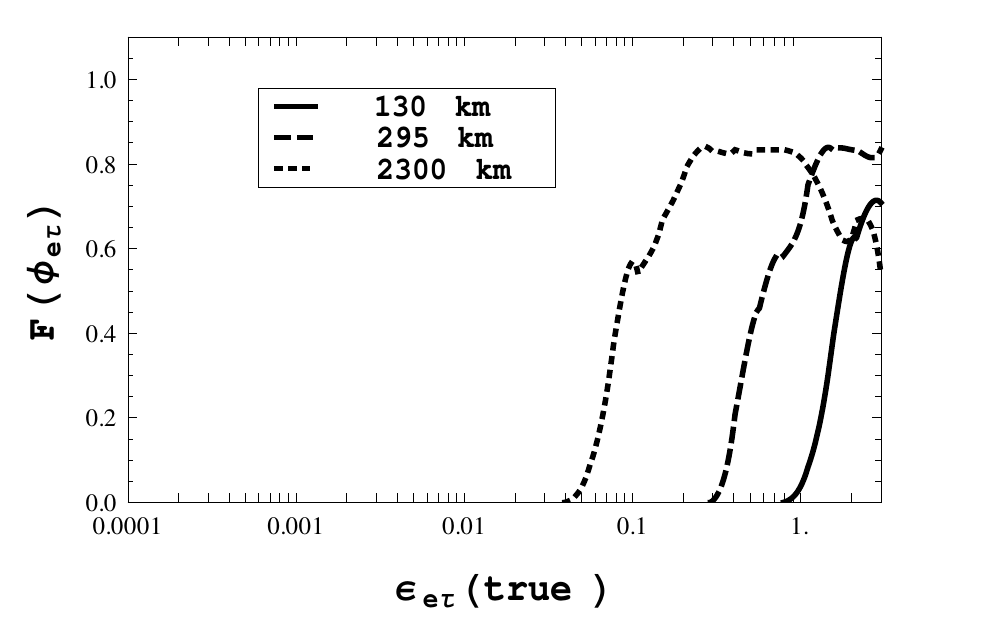}
\end{tabular}
\includegraphics[width=0.45\textwidth]{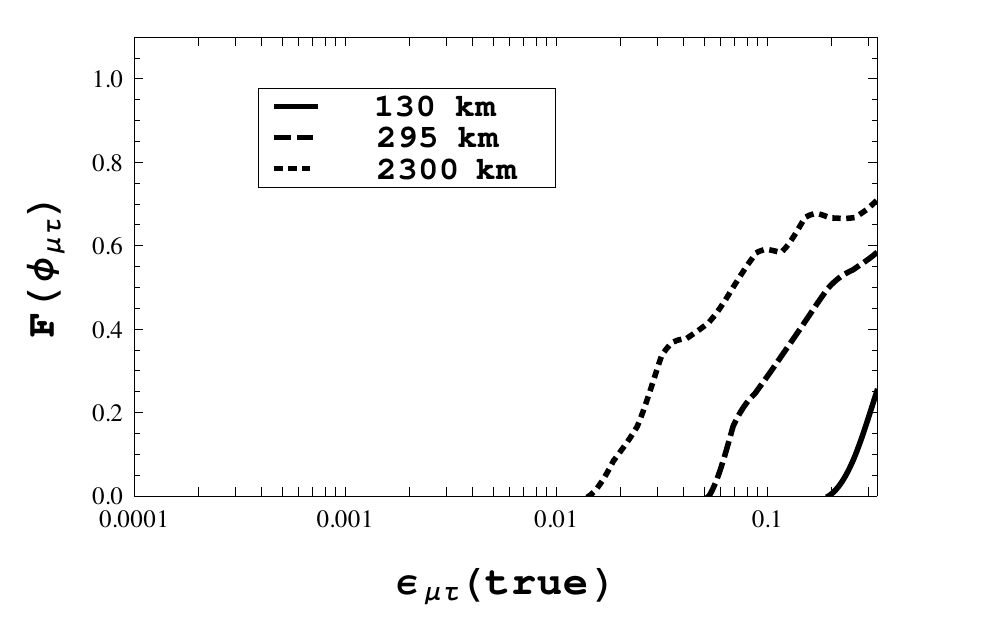}

\caption[] {{\small $F({\phi_{e\mu}})$, $F({\phi_{e\tau}})$ and  $F({\phi_{\mu\tau}})$  (one NSI at a time) for $\delta_{CP}=0$ at 3$\sigma$ confidence level for three different experimental set-up. }}
\label{fig:sbnsid0}
\end{figure*}

\begin{figure}[htb]
\centering
\begin{tabular}{ccc}
\includegraphics[width=0.35\textwidth]{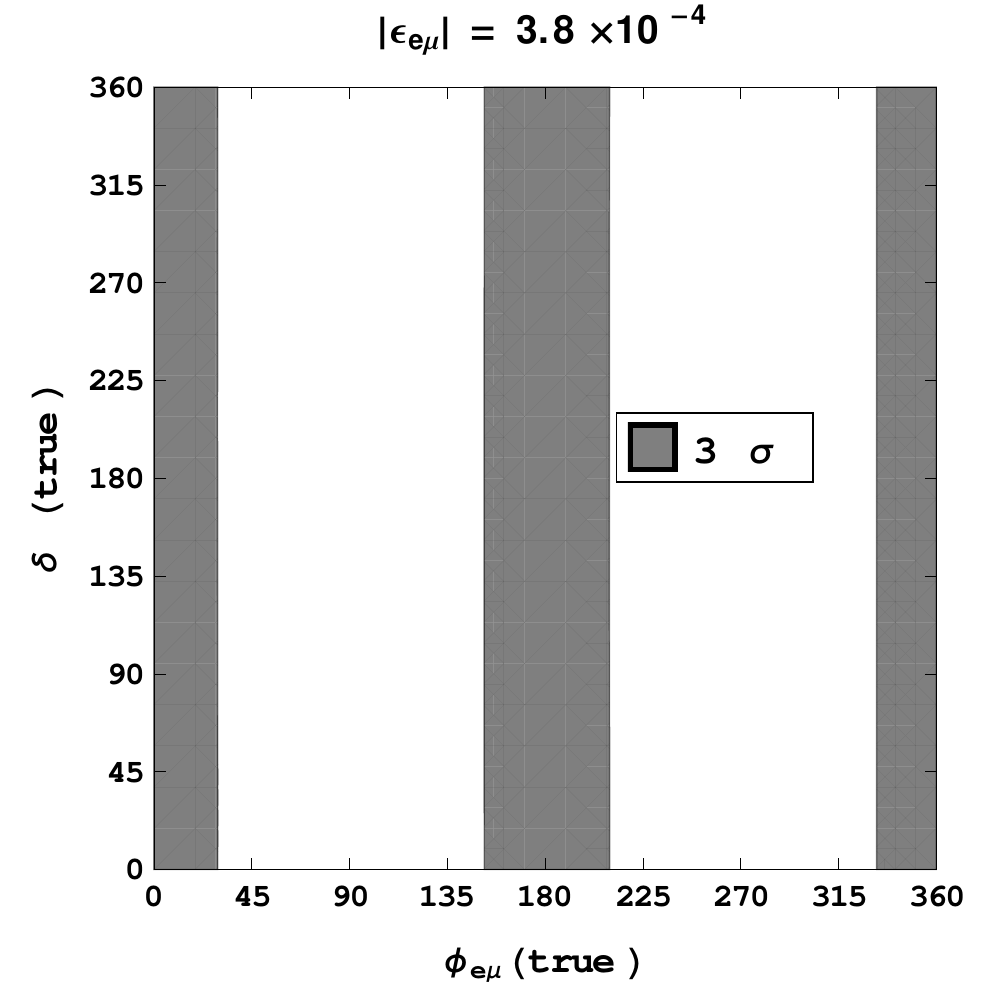}&
\includegraphics[width=0.35\textwidth]{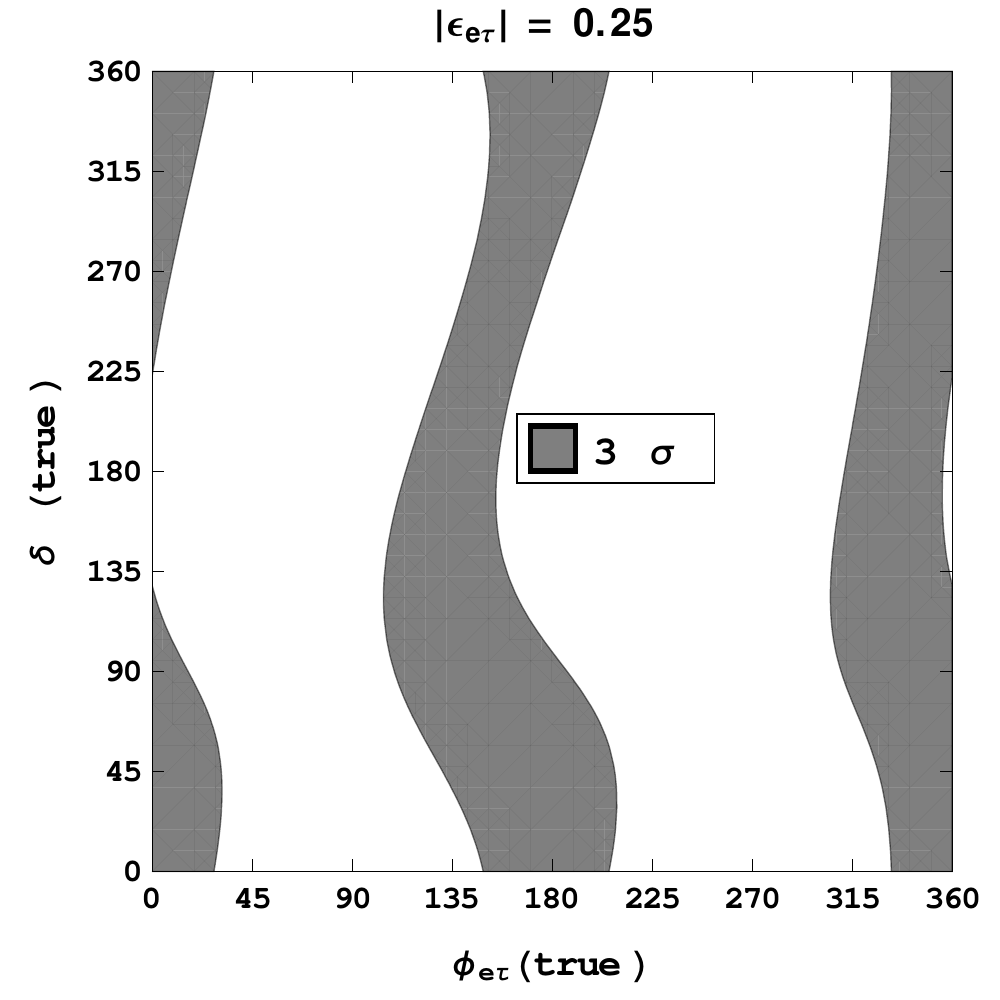}&
\includegraphics[width=0.35\textwidth]{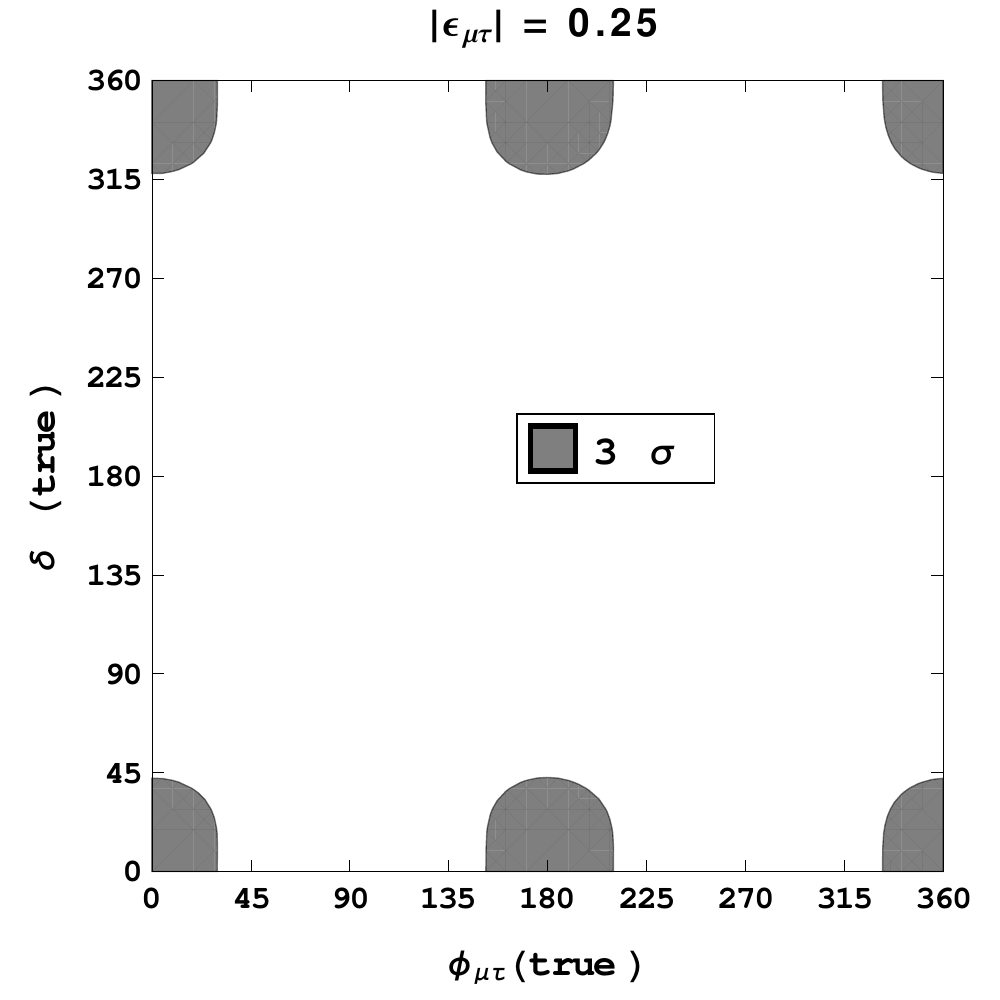}\\
\includegraphics[width=0.35\textwidth]{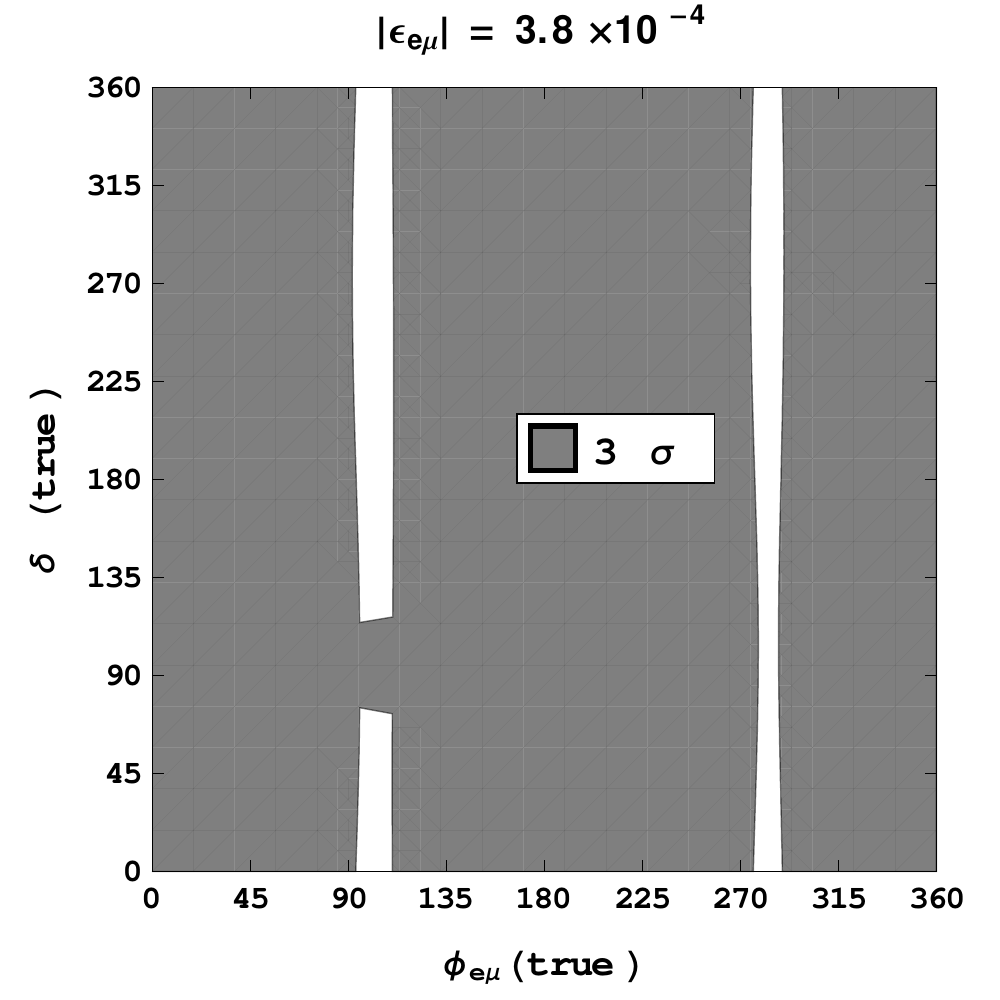}&
\includegraphics[width=0.35\textwidth]{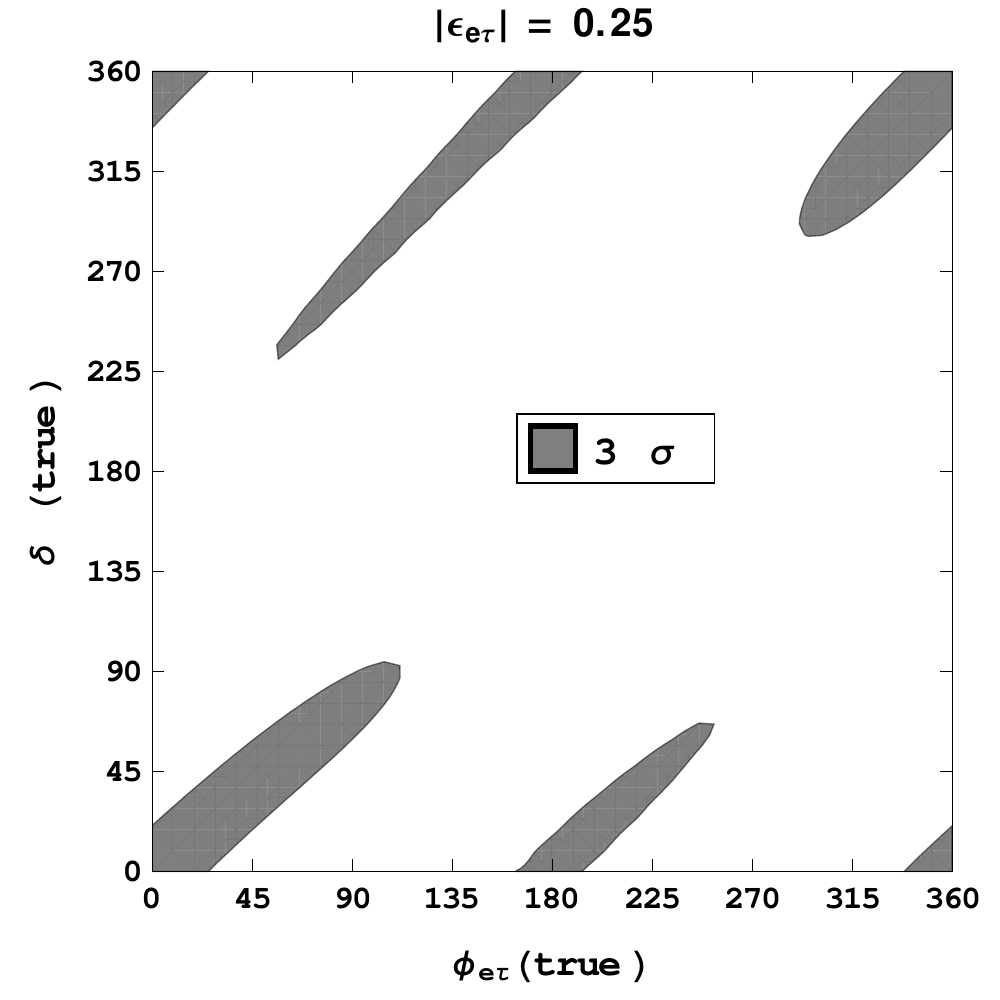}&
\includegraphics[width=0.35\textwidth]{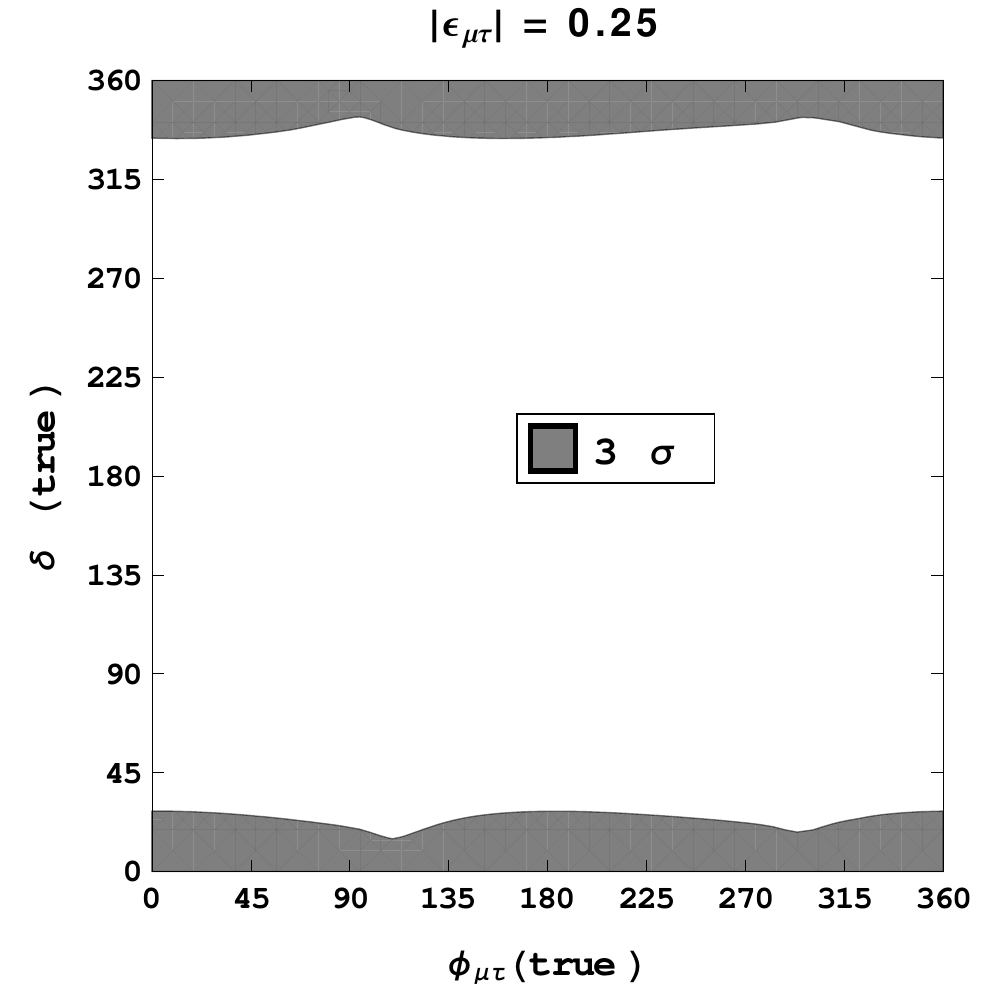}
\end{tabular}
\caption[] {{\small  Allowed region (unshaded) for total $CP$ violation discovery reach $F(\delta)$ for two different baselines T2HK (top panel) and CERN-Pyh\"asalmi set-up (bottom panel)  considering NSI phases $\phi_{e\mu}$, $\phi_{e\tau}$ and $\phi_{\mu\tau}$. 
}}
\label{fig:sbnsiphase}
\end{figure}
  
Here we first discuss the $CP$ violation discovery reach without any NSI for the three experimental set-ups (a), (b) and (c). 
Then we make a comparative study between no NSI and with real NSI. We also consider off-diagonal NSIs with phases. 
The Figures are given only for normal hierarchy (true). However,  there is no significant departures in the results for inverted  hierarchy from that
with normal hierarchy. 

In Figure \ref{fig:delta1} we have shown the $\Delta \chi^2$ values versus $\delta$(true) for SM interactions of neutrinos with matter from which the discovery 
reach can be obtained at different confidence levels. We have fixed $\delta$(test) to 0 and $\pi$ and have marginalized over hierarchy for every  $\delta$(true) value. To calculate $CP$ fraction from Figure  \ref{fig:delta1} at $3 \sigma $ one is required to find the fraction of the total allowed range of $\delta $ for which  $\Delta \chi^2$ values are above the solid horizontal line in Figure \ref{fig:delta1}. Particularly at $3 \sigma $ confidence level  for CERN-Fr\'ejus set-up the $CP$ violation could be discovered 
with $F(\delta)$  of around 0.58 of the possible $\delta$ values for  normal hierarchy (true value) whereas for 
T2HK set-up  these values are around 0.67 and for CERN-Pyh\"asalmi set-up these values are about 0.21.  For longer baseline for normal hierarchy the discovery reach is little bit better than the inverted hierarchy (not shown in the Figure). The discovery reach for longer baseline of 2300 km  was shown earlier by Coloma {\it et al}  \cite{Coloma:2012ma}. So with only SM interactions of neutrinos with matter the short baseline like T2HK set-up seems to be better for good discovery reach of $CP$ violation. This was observed earlier by different authors \cite{Agarwalla:2011hh,others4,Coloma:2012ma,Coloma:2012ut,betabeam}.

\paragraph{\bf Possibilities of Discovery  of $CP$ violation due to $\delta$ for real NSI:} 

If we compare the $F (\delta)$ obtained from Figure \ref{fig:delta1} at $3 \sigma $ with that shown in Figure \ref{fig:delta2} at $3 \sigma $ we find that for smaller values of NSIs satisfying  model dependent bound  for short baseline of CERN-Fr\'ejus  and T2HK set-up , there is
insignificant  effect of NSI on the the discovery reach of $CP$ violation due to $\delta$. However, for CERN-Pyh\"asalmi set-up even
for smaller NSI values one finds the effect on $CP$ violation discovery reach. 
One interesting feature is found from these figures which is that if  one does not observe $CP$ violation due to $\delta $ in shorter baselines (say CERN-Fr\'ejus set-up) there is still a possibility of observing  $CP$ violation
in longer baselines (see CERN-Pyh\"asalmi set-up with off-diagonal NSI like  $\ve_{e\mu}$ and $\ve_{e\tau}$) which could be the signal of NSI with significant strength.  In Figure \ref{fig:delta2} one can see that except $\ve_{\mu\mu}$ for other NSIs the $F(\delta )$ changes considerably, with increase 
in the values  of different NSIs for CERN-Pyh\"asalmi set-up.

\paragraph{\bf Possibilities of Discovery of $CP$ violation due to  complex NSI phases :}

There could be $CP$ violation only due to off-diagonal NSI phases for which $\delta =0$. Such cases have been shown in Figures 
\ref{fig:sbnsid0}.   We observe that for certain NSI value  the $CP$ fraction for longer baselines is more in comparison to that for the shorter baselines. With the increase of $|\varepsilon_{\alpha \beta}|$ there
is increase in discovery of $CP$ fraction in general. If $CP$ violation is not observed in short baselines (say CERN-Fr\'ejus) but it is observed at relatively longer baselines (say CERN-Pyh\"asalmi set-up) then it would signal the presence of NSI  as can be seen from  Figure \ref{fig:sbnsid0}.

\paragraph{\bf Possibilities of Discovery of $CP$ violation due to both $\delta$ and complex NSI phases :}

In Figure \ref{fig:sbnsiphase} in the un-shaded region  discovery of total $CP$ violation could be possible at $3 \sigma$ confidence level. 
Here, we are assuming that nature has two sources of $CP$ violation - one due to $\delta $ phase and other due to one NSI phase. For this we ask whether there is ($\delta - \phi_{ij}$) parameter space where both can be distinguishable from their $CP$ conserving points as discussed earlier.
From the Figure it is seen that  for  T2HK set-up there is more allowed region and as such better discovery reach in the $\delta - \phi_{ij}$ parameter space particularly for NSI - $\ve_{e\mu}$ in comparison to CERN-Pyh\"asalmi set-up .  Particularly for NSI - $\ve_{e\tau}$ and $\ve_{\mu\tau}$, the discovery reach of total $CP$ violation is better for both longer and shorter baselines (like CERN-Pyh\"asalmi set-up and T2HK set-up). Simultaneous consideration of experimental data from both these set-ups  could improve the discovery reach further because of presence of non-overlapping un-shaded discovery regions in the 
upper panel two right Figures and lower panel two right Figures corresponding to two different NSIs. 

From Figure  \ref{fig:delta1} it is observed that for certain range of values of $\delta $ (about $0^\circ \mbox{to} \;25^\circ$, $160^\circ \mbox{to} \;208^\circ$, $340^\circ \mbox{to} \; 360^\circ$), the $CP$ violation due to $\delta $  could remain unobservable.
However, even if nature has such values of $\delta $,  from Figure \ref{fig:sbnsiphase} one can see that   total $CP$ violation could be observed in such cases  in presence of 
NSI for a wide range of values of different NSI phases. However, disentangling the sources of two types of $CP$ violating phases (one from NSI phase and other from $\delta$) could be difficult. Here the total $CP$ fraction can be thought of as the ratio of the un-shaded region divided by the total region covered by $\delta - \phi_{ij}$ parameter space. As for example, in the upper panel extreme left Figure
such total $CP$ fraction is about $75 \%$. For both the set-ups for $\ve_{e\tau}$ such $CP$ fraction is even more. For CERN-Pyh\"asalmi set-up for $\ve_{e\mu}$ the total $CP$ fraction is not good.


\subsection{\bf Neutrino factory}
\label{sec:nu}
We discuss below the experimental set-up and systematic errors for neutrino factory.  We have presented our results for discovery 
reach of $CP$ violation 
for different baselines of length 730 km  (FNAL-Soudan),1290 km (FNAL-Homestake) and 1500 km (FNAL-Henderson) due to $\delta $ and other off-diagonal NSI phases $\phi_{ij}$. We have shown the effect of the absolute values of different $\varepsilon_{\a\b}$ in that.
For this analysis we have considered one NSI at a time.

\subsubsection{\bf Experimental set-ups and systemetic errors}
We have used a large magnetised iron neutrino detector(MIND) \cite{mind1} with a toroidal magnetic field having a mass of 100 KTon. MIND can also be described as an iron-scintillator calorimeter. This detector has the capability of excellent reconstruction and
charge detection efficiency. In this section we have considered muons in a storage ring consisting of both $\mu^+$ and $\mu^-$  which decay with  energies of 10 GeV. We consider $5\times 10^{21}$ stored muons. The golden channel ($\nu_e \rightarrow \nu_{\mu}$ oscillation channel) where the charged current interactions of the $\nu_{\mu}$ produce muons of the opposite
charge to those stored in the storage ring (generally known as wrong-sign muons), is the most promising channel to explore $CP$ violation at a neutrino factory. The detector that we are considering in this work - MIND is optimized to exploit the golden channel
oscillation as this detector has the capacity to  easily identify signal i.e. a muon with a sign opposite to that in the muon storage ring.   
 Different 
oscillation channels which have been considered as signals and backgrounds \cite{mind1} in the analysis are 
shown in table \ref{table:channelnufact}.   
We have taken the migration matrices for the true and reconstructed neutrino energies  as given in reference \cite{mind1}. The signal  and background efficiencies are taken into account in those  matrices. We have taken into account
the reconstruction of $\tau$ contamination coming from the \emph{silver channel} $\nu_e \rightarrow \nu_\tau$ as background.
 We have considered systematic errors to be $1\%$. In this work we have considered a running time of 5 years for both $\mu^+$ and $\mu^-$ .
\begin{table}[ht]
\begin{tabular}{|l|l|l|l|}
\hline
 & Channel Name & $\mu^+$ & $\mu^-$ \\ \hline
 \multirow{1}{*}{Signal} & Golden Channel & $\nu_e \rightarrow \nu_\mu$ & $\overline{\nu}_e \rightarrow \overline{\nu}_\mu$ \\   \hline
 \multirow{5}{*}{Background} & $\nu_e$ disappearance channel & $\nu_\mu \rightarrow \nu_e$ & $\overline{\nu}_\mu \rightarrow \overline{\nu}_{e}$ \\
& Silver Channel & $\nu_e \rightarrow \nu_\tau$ & $\overline{\nu}_e \rightarrow 
 \overline{\nu}_\tau$ \\
& $\nu_\mu$ disappearance channel & $\overline{\nu}_\mu \rightarrow 
 \overline{\nu}_\mu$ & $\nu_\mu \rightarrow \nu_\mu$ \\
 & Platinum Channel & $\overline{\nu}_\mu \rightarrow 
 \overline{\nu}_e$& $\nu_\mu \rightarrow \nu_e$ \\
 & Dominant Channel& $\overline{\nu}_\mu \rightarrow 
 \overline{\nu}_\tau$& $\nu_\mu \rightarrow \nu_\tau$\\
 \hline
\end{tabular}
\caption{ Different oscillation channels considered as signals and backgrounds in the analysis.}
\label{table:channelnufact} 
\end{table}

\begin{table*}
\begin{tabular}{|c|c|c|c|}
\hline
\hline
 &730 km & 1290 km  & 1500 km \\
 \hline
SM &  248491 & 222097 & 210819\\
\hline
$\varepsilon_{ee}$ &  248715 & 222644 & 211527\\
\hline
$\varepsilon_{e\mu}$& 316049 & 283709& 269412 \\
\hline
$\varepsilon_{e\tau}$ & 251331 & 229148 & 220134 \\
\hline
$\varepsilon_{\mu\tau}$& 248049 & 220985 & 209394\\
\hline
$\varepsilon_{\mu\mu}$& 248682 & 222667 & 211597 \\
\hline
$\varepsilon_{\tau\tau}$& 248072 & 220964 & 209313\\
\hline
 \end{tabular}
\caption{ Number of  ${\nu_{e}\rightarrow\nu_\mu}$ events for 730 km  (FNAL-Soudan),1290 km (FNAL-Homestake) and 1500 km (FNAL-Henderson) for SM and for different NSI (one at a time each of which equals to $\a$).}
\label{table:eventnufact} 
\end{table*}
In table \ref{table:eventnufact} for $\delta =0$ we have shown the expected number of events for three baselines in neutrino factory, for no NSI and also for different
real NSI , each of which equal to $\a$ (one NSI at a time). We have considered the central values of various parameters
as shown in table \ref{table:mix} and have used PREM profile \cite{prem} for matter densities. Like superbeam case here also for neutrino factory one can see that there is significant effect of  $\ve_{e\mu}$  and $\ve_{e\tau}$ on the number of $\nu_e \rightarrow  \nu_{\mu}$ events.  
Possibilities of significant effect of these NSIs can be seen in the expression of oscillation probabilities ${\nu_{e}\rightarrow\nu_\mu}$ (which can be obtained using Eq.\eqref{nufacttosb} in   Eq.\eqref{eq:pro2}).

\subsubsection{\bf Results}
\begin{figure}[ht]
\centering
\includegraphics[width=0.7\textwidth]{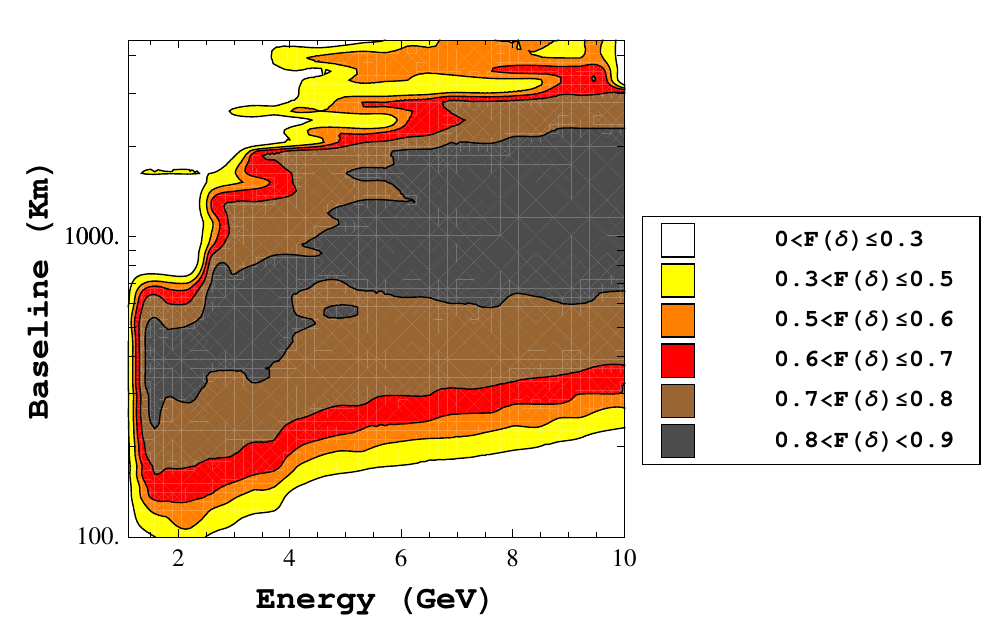}
\caption[] {{\small $F(\delta)$ with only SM interactions of neutrinos with matter for baselines of different lengths and different muon energies $E_\mu$ at  5$\sigma$ confidence level.}}
\label{fig:elsm}
\end{figure}
The optimization of $CP$ violation discovery reach has been done earlier \cite{Ballett:2012rz,Agarwalla:2010hk} for different baselines and different parent muon energy  when only  SM interactions of neutrinos with matter during propagation is present. 
 We have re-analysed this optimization based on the updated MIND detector characteristics 
and the migration matrices as given in \cite{mind1}.  We present the result for neutrino factory at $5 \sigma $ confidence level unlike superbeam case where we have given the results at $3 \sigma $ confidence level. This is because MIND detector for neutrino factory is found to give quite good $CP$ violation discovery reach. It is found that at 5$\sigma$ confidence level the $CP$ fraction (shown as $F(\delta)$
in Figure \ref{fig:elsm})
 of about ($0.8 \leq F(\delta) < 0.9$ ) is possible  for baselines ranging from about 300 to 800 km  for energies  lesser than 5 GeV and for baselines ranging from about 700 to 2000 km for energies 6-10 GeV respectively.  Based
on high $CP$ fraction discovery potential as found in this Figure we have chosen 10 GeV muon energy 
and a few baselines  which are : 730 km(FNAL-Soudan), 
1290 km (FNAL-Homestake) and 1500 km (FNAL-Henderson). 

\begin{figure*}[h!]
\centering
\begin{tabular}{cc}
\includegraphics[width=0.45\textwidth]{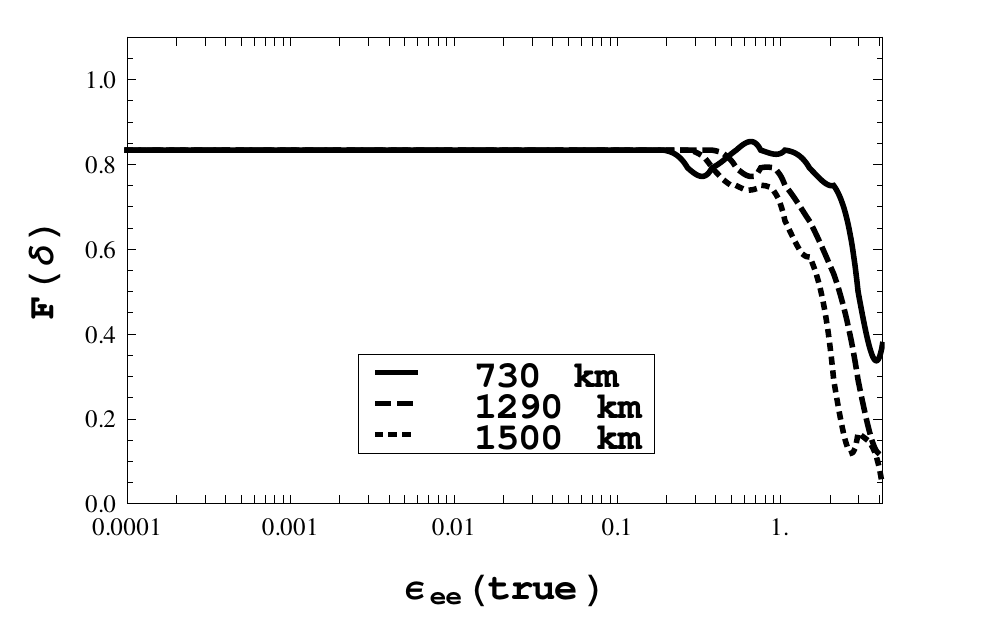}&
\includegraphics[width=0.45\textwidth]{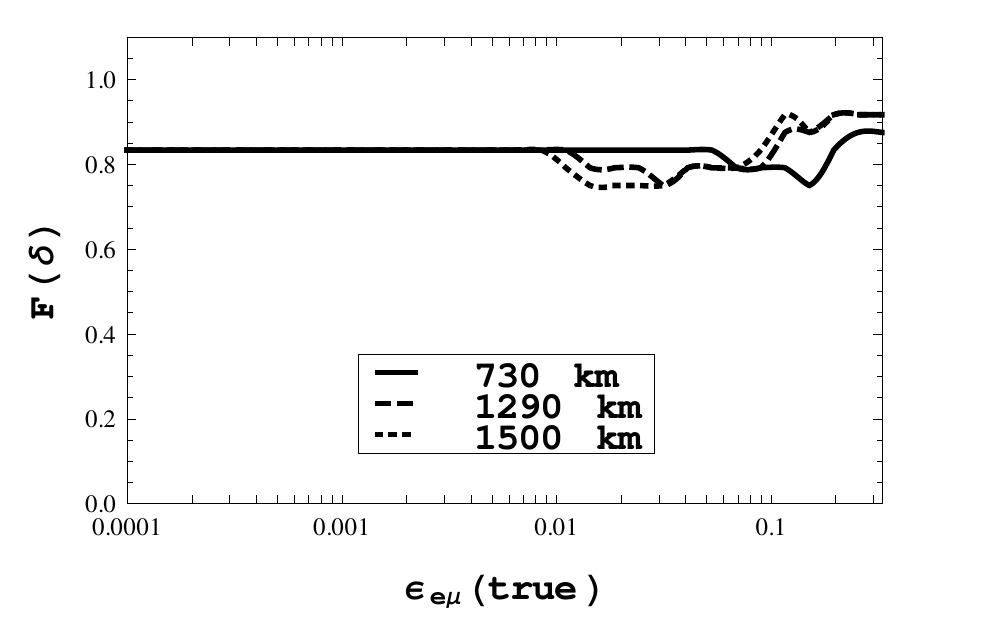}\\
\includegraphics[width=0.45\textwidth]{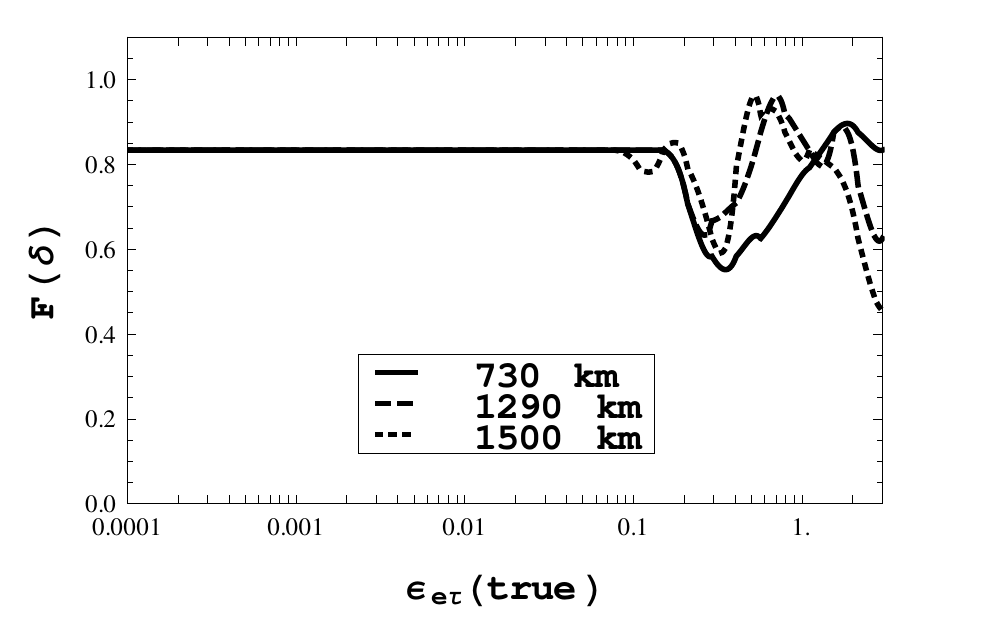}&
\includegraphics[width=0.45\textwidth]{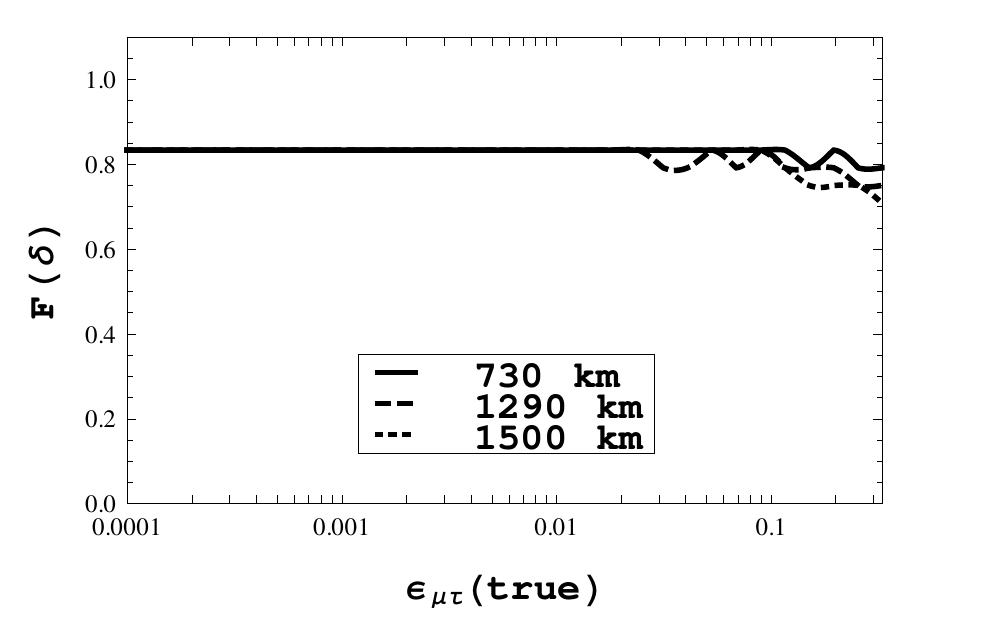}\\
\includegraphics[width=0.45\textwidth]{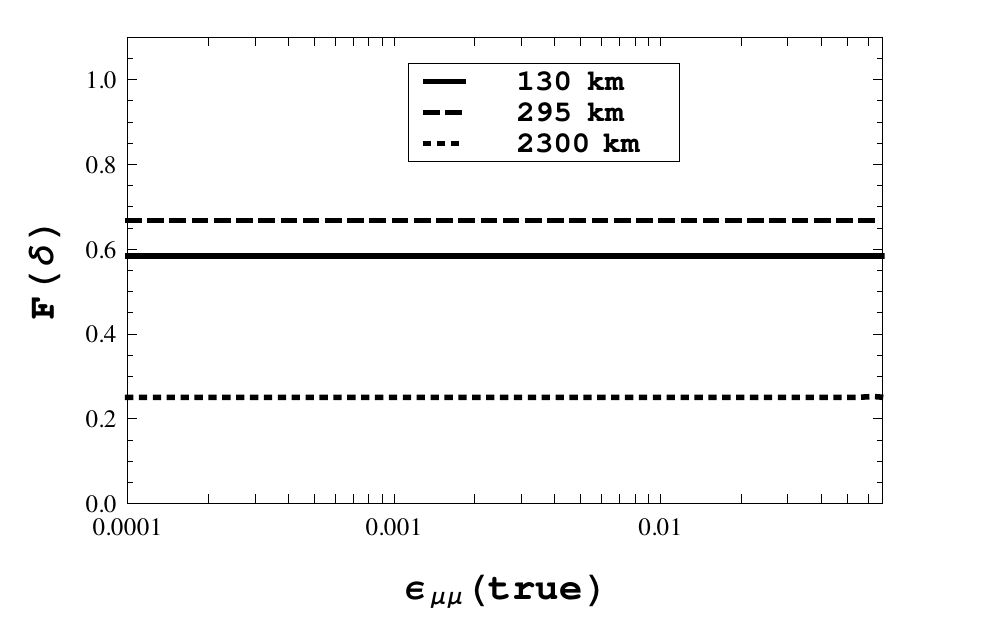}&
\includegraphics[width=0.45\textwidth]{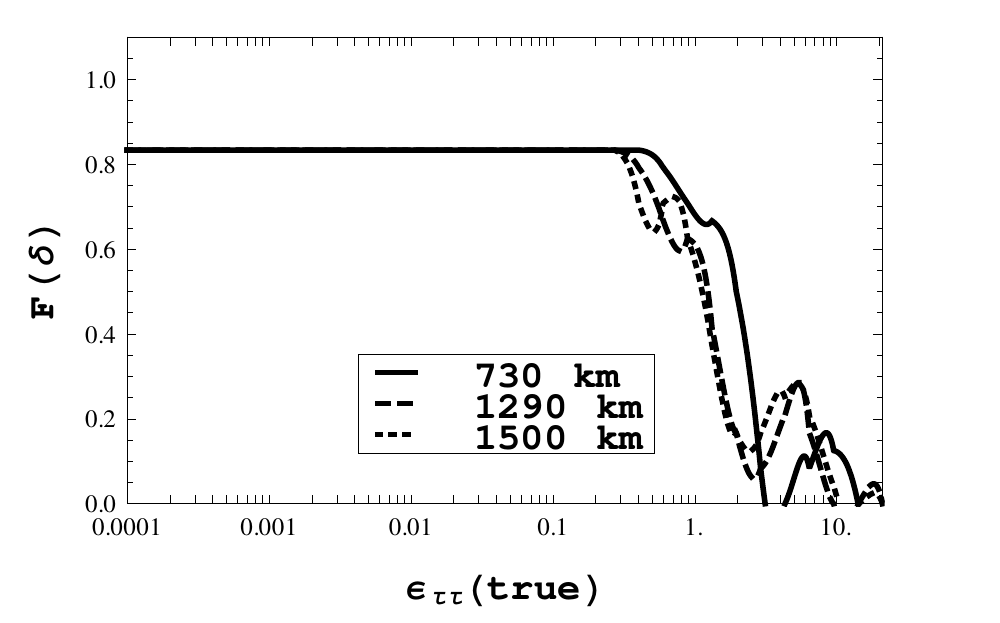}
\end{tabular}
\caption[] {{\small $F(\delta)$ versus real NSI ($\varepsilon_{ij}$) at $5\sigma$ confidence levels.}}
\label{fig:cpfrac}
\end{figure*}

\begin{figure*}[h!]
\centering
\begin{tabular}{cc}
\includegraphics[width=0.45\textwidth]{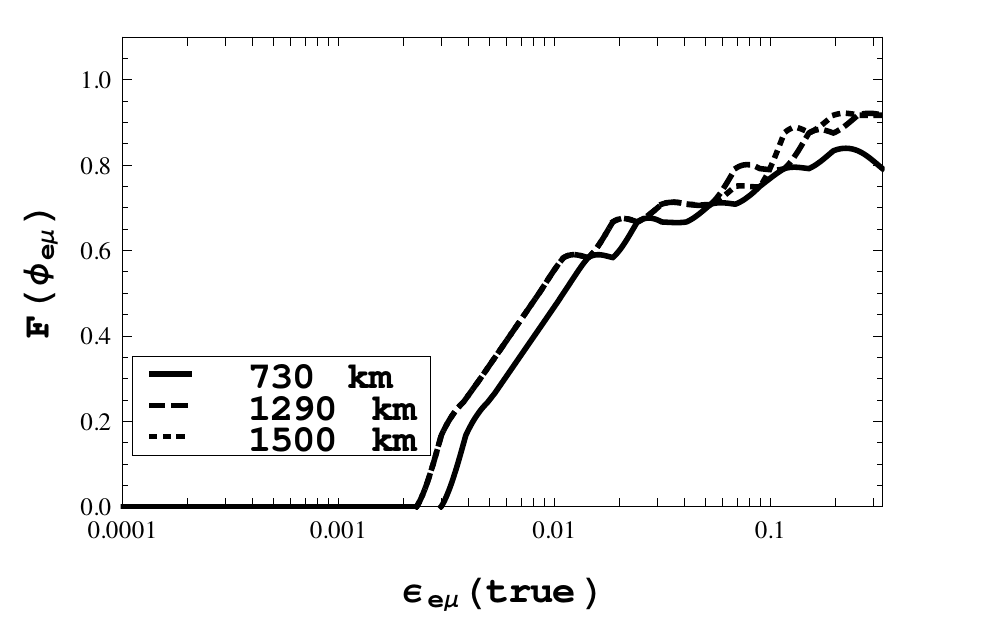}&
\includegraphics[width=0.45\textwidth]{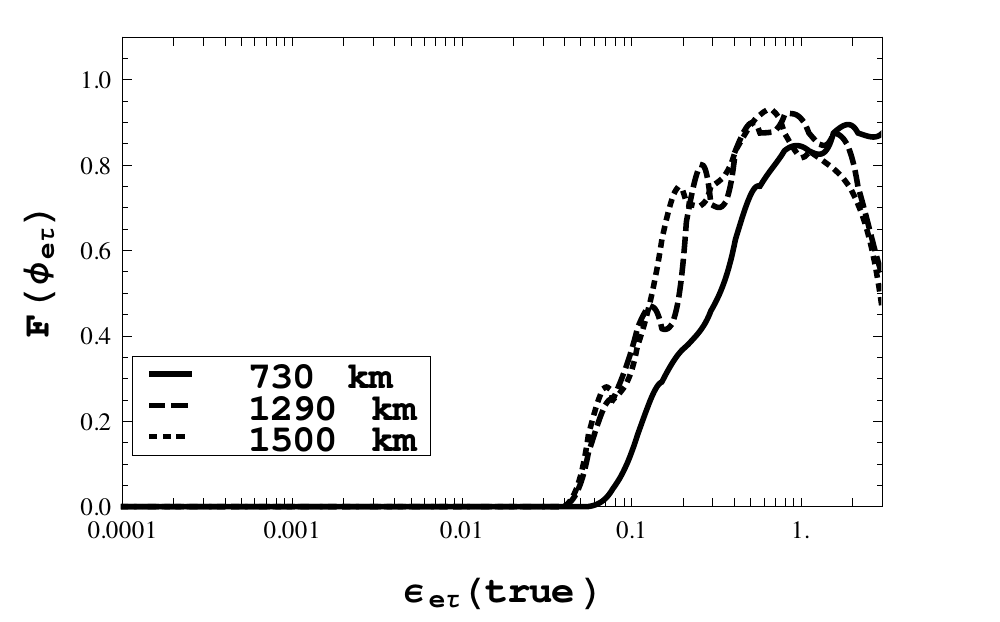}
\end{tabular}
\includegraphics[width=0.45\textwidth]{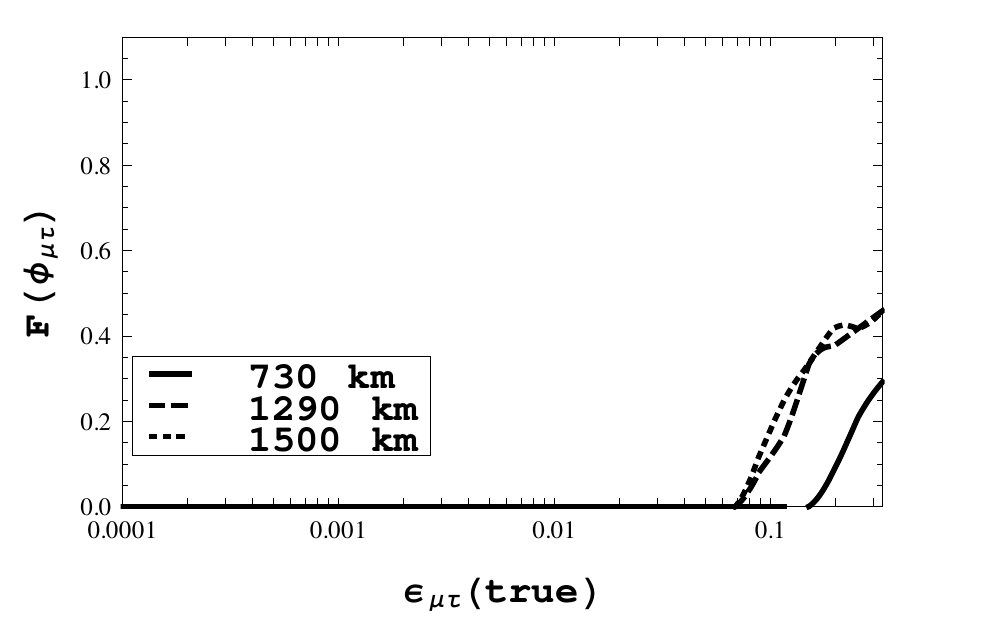}
\caption[] {{\small $F({\phi_{e\mu}})$, $F({\phi_{e\tau}})$ and  $F({\phi_{\mu\tau}})$   for $\delta_{CP}=0$ at 5$\sigma$ confidence level for three baselines. }}
\label{fig:nsid0}
\end{figure*}

Next we have tried to answer the question that had there been non-Standard 
interactions what could have been their effect on the $CP$ violation discovery reach for such experimental set-ups. 
While taking into account NSI effect, for off-diagonal NSI we have also considered the effect of NSI phases over $\delta_{CP}$ violation.   
In contrast to the results on superbeam we have presented results for neutrino factory at $5\sigma$ confidence level in Figures \ref{fig:cpfrac} and \ref{fig:nsid0} because in most of 
the cases the discovery reach of $CP$ violation is far better in neutrino factory set-ups than those in superbeam set-ups particularly
for NSIs $\lsim 1$. However, for
comparison of the results of neutrino factory set-ups with that of superbeam set-ups we have considered the $3 \sigma $ results for
both type of set-ups. For brevity we have avoided presenting the results at  $3 \sigma $  
corresponding to  Figures \ref{fig:cpfrac} and \ref{fig:nsid0}.

\paragraph{\bf Possibilities of Discovery of $CP$ violation due to $\delta$ for real NSI:} 
In  
Figure \ref{fig:cpfrac}, we have studied $F(\delta)$ in the presence of real NSI (NSI phases have been chosen to be zero) for different baselines of length 730 km, 1290 km and 1500 km at $5 \sigma$ confidence level. Here, in plotting the Figures we have considered the NSI values upto the model independent  bounds as shown in table \ref{table:bound}. For lower values of NSI there is essentially negligible effect on discovery reach of $CP$ violation which is seen in the Figure as horizontal straight line. This corresponds  to $F(\delta)$ due to SM interactions only which can be verified from Figure \ref{fig:elsm} at muon energy of 10 GeV
for the appropriate baseline.   For  NSI $\varepsilon_{\mu\mu}$ with model dependent bound  there is  insignificant effect on $F(\delta)$      
whereas for  other NSIs  there are some  effect on $F(\delta)$.  Except $\ve_{e\mu}, \; \ve_{e\tau}$ and $\ve_{\mu\mu}$
for other NSIs there is considerable decrease in discovery reach of $CP$ violation with the increase of NSI values.
Even if there is $CP$ violating phase $\delta$ but as one
can see that in presence of higher allowed values of NSI - $\varepsilon_{\tau\tau} $ there could be no discovery of $\delta_{CP}$  
violation. 

The results are shown in Figure \ref{fig:cpfrac} at $5 \sigma $ confidence level. However, at $ 3 \sigma $ also the decreasing features 
of $\ve_{ee}$ and $\ve_{\tau\tau}$  $\gsim 1$  is present and  in comparison to Figure \ref{fig:delta2} it is found that for large NSI values of $\ve_{ee},\ve_{\tau \tau} \gsim 1$ superbeam T2HK set-up is found to be better for the discovery reach of $CP$ violation as compared to neutrino factory set-ups.
But for lower values of NSIs  particularly $\ve \lesssim 10^{-1}$ the neutrino factory set-ups are found to be much better for the discovery of 
$CP$ violation at $3 \sigma $ and even at $5 \sigma $ the $CP$ fraction is more in neutrino factory set-ups.

\paragraph{\bf Possibilities of Discovery of $CP$ violation due to complex NSI phases :}
In Figure \ref{fig:nsid0} we have addressed the question of what could be 
the $CP$ fraction for discovery of $CP$ violation if Dirac phase $\delta$ 
is absent in PMNS mixing matrix and $CP$ violation comes from purely NSI phases.
  We observe that for certain NSI value  the $CP$ fraction for longer baselines could be more in comparison to that for the shorter baselines. With the increase of $|\varepsilon_{\alpha \beta}|$ there
is increase in discovery of $CP$ fraction in general. If $CP$ violation is not observed in short baselines (say 730 km) but it is observed at relatively longer baselines (say 1500 km) then it would signal the presence of NSI  as can be seen from our Figures although for
$\varepsilon_{e \mu}$ it is difficult to distinguish the effect between short and long baselines considered by us. Comparing $3\sigma $ result of neutrino factory (not shown here) with superbeam result of Figure  \ref{fig:sbnsid0} it is found that in the presence of $\varepsilon_{e\mu}$ the Neutrino factory set-up gives better 
$CP$ violation discovery reach whereas for $\ve_{\mu \tau}$ the superbeam CERN-Pyh\"asalmi set-up is better . For $\ve_{e\tau}$ the neutrino factory 
set-up is slightly better than superbeam set-up.  
\begin{figure*}[t]
\centering
\begin{tabular}{cc}
\includegraphics[width=0.4\textwidth]{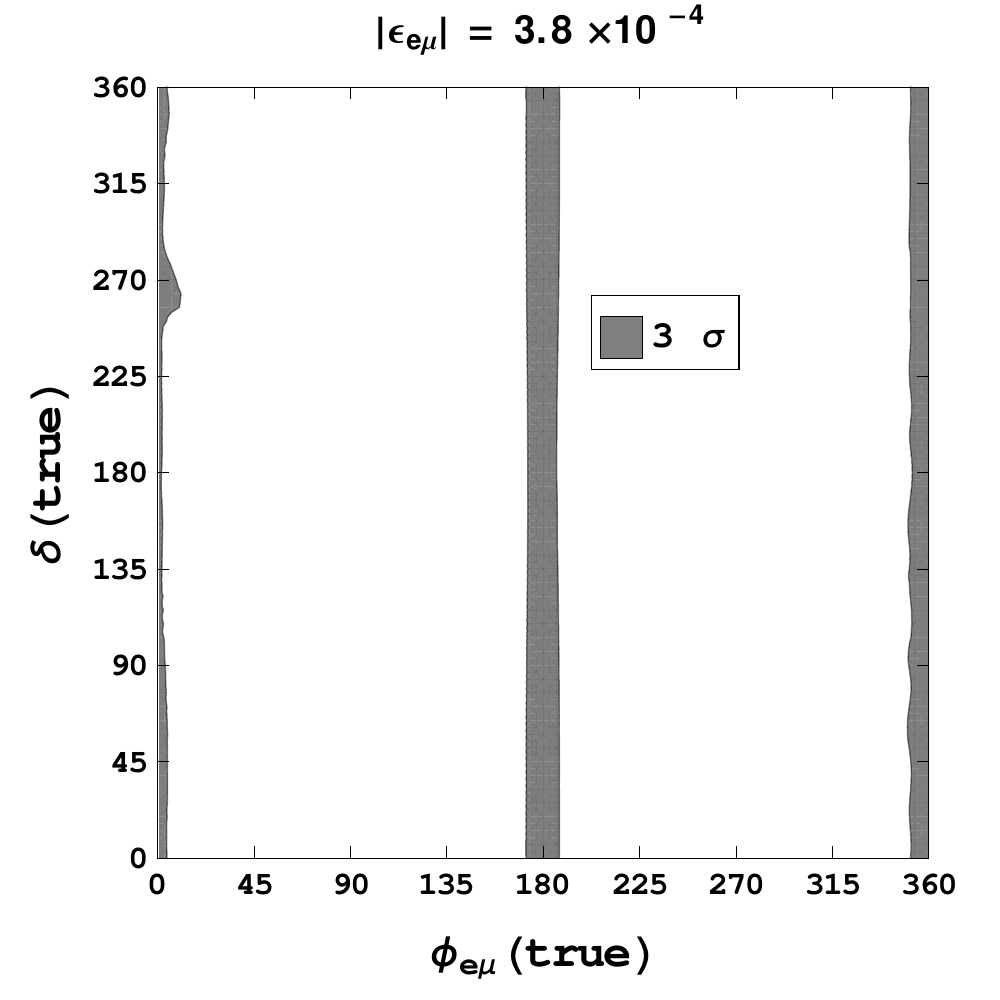}&
\includegraphics[width=0.4\textwidth]{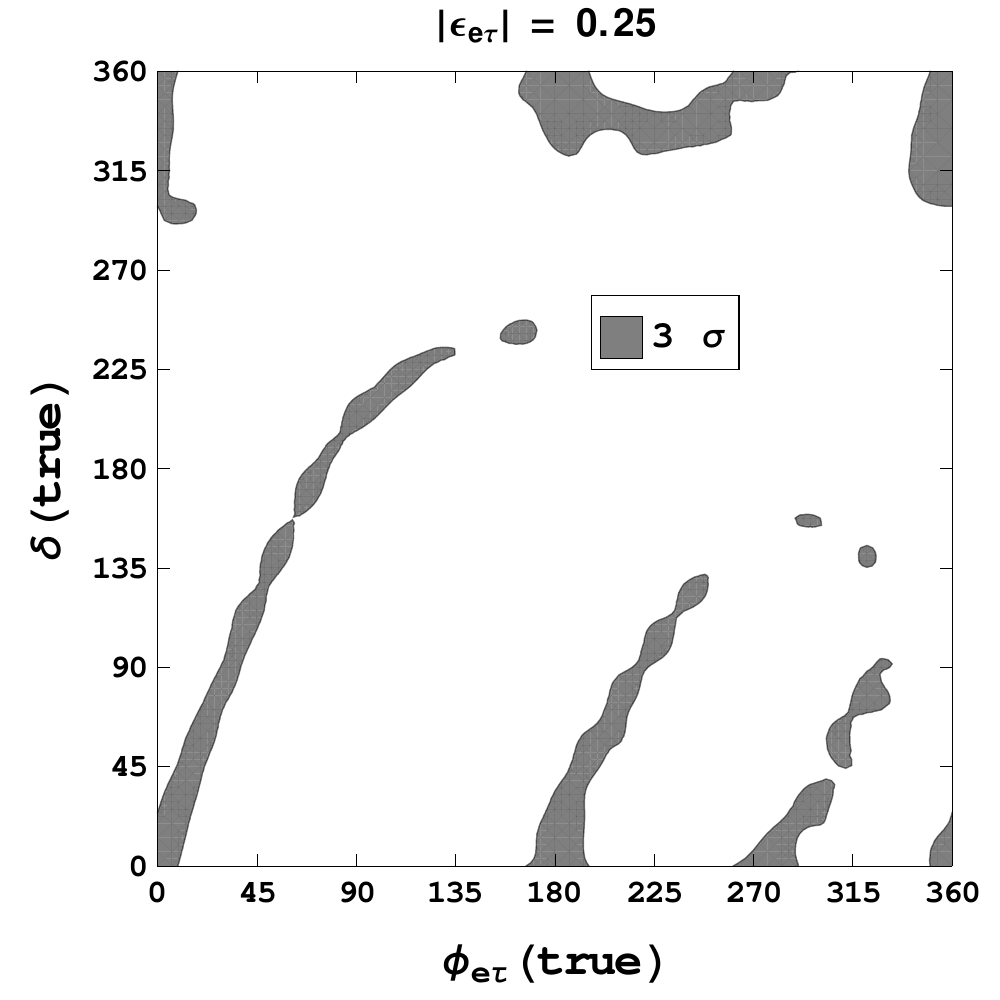}\\
\end{tabular}
\includegraphics[width=0.4\textwidth]{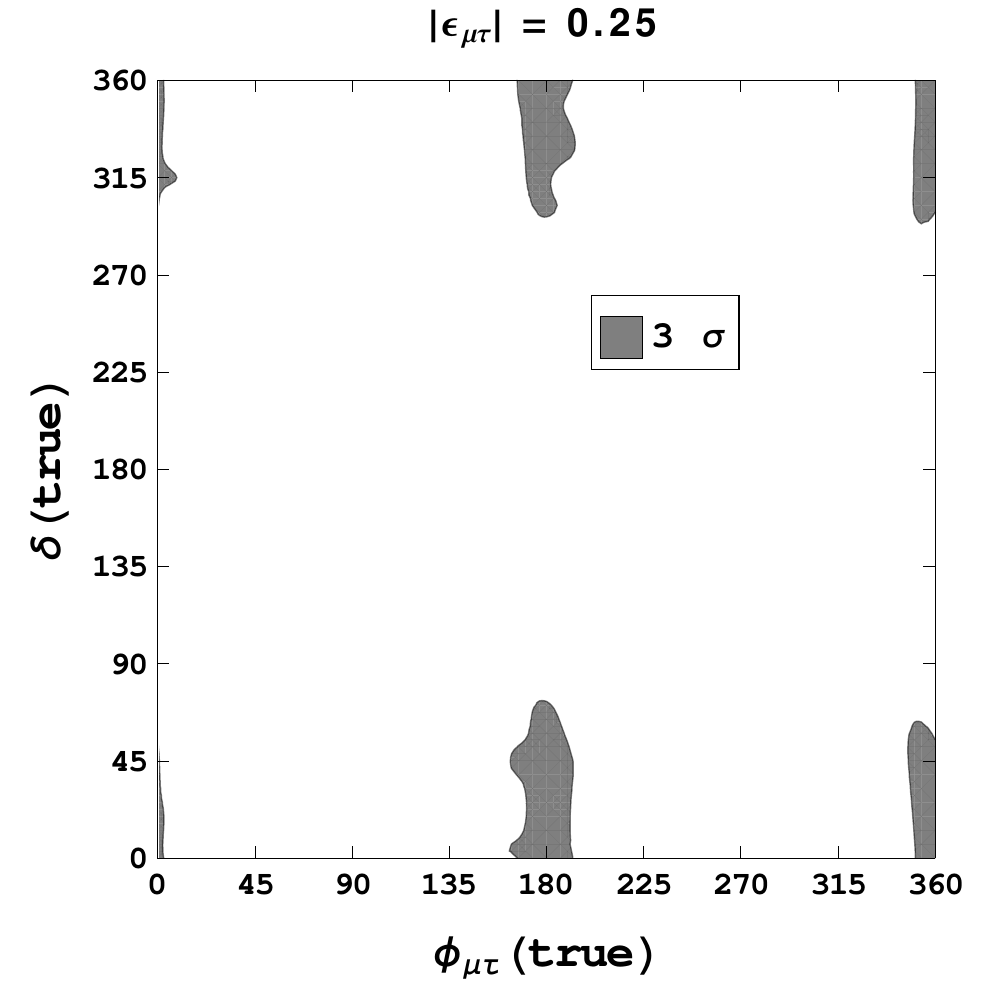}\\
\caption[] {{\small $\delta$ versus phase ($\phi_{ij}$) considering the value of NSIs($\varepsilon_{ij}$) at the uppermost value with model dependent bounds. Unshaded regions correspond to $CP$ violation discovery reach.}}
\label{fig:phcol}
\end{figure*}

\paragraph{\bf Possibilities of Discovery of $CP$ violation due to both $\delta$ and complex NSI phases :}
In Figures \ref{fig:phcol} we have considered the 
case for 730 km baseline where the $CP$ violation might come from $\delta_{CP}$ as well as from  
NSI phase $\phi_{\alpha \beta}$. For 1290 km and 1500 km baselines the essential feature is almost same (not shown in the Figure).
Here we have chosen the uppermost value of NSI with model dependent bound.  
The total $CP$ violation due to $\delta $ and $\phi_{ij}$ could be observable even for those $\delta $ which could remain unobservable 
in absence of NSI phases for some ranges of $\delta $ values.
 In fact, the $CP$ violation discovery reach could be possible
for the whole range of allowed $\delta $ values for some specific values of $\phi_{ij}$ as can be seen from Figures.
But it would be difficult to conclude whether the observed $CP$ violation is due to $\delta $ or any  NSI phase. As compared to
Figure \ref{fig:sbnsiphase} due to different superbeam set-ups, we find better discovery reach of $CP$ violation for 
neutrino factory set-up for 730 km baseline in Figures \ref{fig:phcol}. As for example, in the extreme left Figure for $\ve_{e\mu}$
the total $CP$ fraction is around $94 \%$ for neutrino factory in contrast to $75 \%$ for the T2HK superbeam set-up as shown in Figures \ref{fig:sbnsiphase}. 

Comparing Figures \ref{fig:phcol} and \ref{fig:sbnsiphase} we find that for $\phi_{e\tau}$   if we combine 
experimental data from superbeam set-ups and neutrino factory set-ups, there are scopes to improve the total $CP$ violation discovery 
reach further.



\subsection{\bf monoenergetic neutrino beam}
\label{sec:mono}
We discuss below  the possible  discovery reach of $CP$ violation for two different nuclei, $^{110}_{50}$Sn and $^{152}$Yb 
which is  considered for the electron capture experiment. We also discuss our procedure
of energy resolution at the detector for numerical simulation.   
Earlier in section \ref{sec:prob} we have shown how  $\nu_e \rightarrow \nu_\mu$ oscillation probability depends on $\delta$ for
shorter baseline ($A \sim \alpha $). 
Based on the oscillation probability and its' variation with respect to $\delta $ we   discuss the procedure
for choosing suitable boost factor $\gamma $ numerically for specific  baseline and specific  nuclei considered for the neutrino beam. We mention 
below four experimental set-ups that we have considered for analysis of discovery reach of $CP$ violation with monoenergetic neutrino beam.   

\subsubsection{\bf Suitable boost factor, neutrino energy from $\nu_e \rightarrow \nu_\mu$ oscillation probability}
The most suitable candidate for producing neutrino beam from electron capture process would be the one with a low Q value and high boost factor, $\gamma$ \cite{mono3}. Also it would be preferable if the nuclei has a short half life. The reason for these is as follows. We need neutrino energy around the  peak of the oscillation probability where variation due to $\delta $ is significant and as such $\frac{\Delta m^2_{31}L}{4E} \approx (2 n + 1) \frac{\pi}{2}$. Considering $E=2\gamma Q$, it follows that $\gamma = \frac{\Delta m^2_{31}L}{4\pi Q}$ as for example, for the first oscillation peak. For sufficiently high $\gamma$ almost all neutrinos are expected to go through the detector. Then to satisfy the above condition we   need to lower Q value. Then another condition is,  $\gamma \tau < T$ where $T$ is the time considered to perform the experiment so that all the nuclei decay and
$\tau $ is the half life of the nuclei. If $\gamma$ is increased then the  half life $\tau$ is required to be small. So the  preferable factors considered in choosing the candidate for producing neutrino beam from electron capture process are -  low Q value,  small half life $\tau$ and  high $\gamma$ \cite{mono3}. Although higher $\gamma $ needs technological advancement for the accelerator.

The isotope, $^{110}_{50}$Sn, has
Q = 267 keV in the rest frame and a half life of 4.11 h. As it
has a low Q value so one may consider high $\gamma$ value. However, it has a longer half life as compared to other nuclei like $^{150}Er$, $^{152}$Yb, $^{156}$Yb, $^{150}$Dy, $^{148}$Dy \cite{mono2,mono11} whose half lifes are 18.5 seconds, 3.04 seconds, 261 seconds, 7.2 min and 3.1 min respectively. However, these nuclei have larger Q values of the order of $10^3$ keV. On the other hand, considering effective running time
per year as $10^7$  second all the nuclei for isotope $^{110}_{50}$Sn will not decay. But for $\gamma =500$ or 320  (as considered in our analysis to  obtain the suitable  neutrino energy $E$ 
resulting in high oscillation probability)  the 
useful decays are respectively about 0.608 or 0.768 times the total number of $^{110}_{50}$Sn nuclei considered. So there is not much suppression in numbers of nuclei. Hence although $^{110}_{50}$Sn has a larger half life, due to its lower Q value there is scope to consider higher $\gamma$
for CERN-Fr\'ejus or 250 km baselines. For these reasons we have preferred isotope, $^{110}_{50}$Sn in comparison to other nuclei. However, there is recent study on finding suitable candidate nuclei for electron capture process and it has been found that $^{150}Er$, $^{152}$Yb, $^{156}$Yb nuclei have dominant electron capture decay
 to one level. Particularly, $^{152}$Yb has been found to be most suitable one \cite{mono11}.
For this reason,  apart from nuclei $^{110}_{50}$Sn we shall consider $^{152}$Yb also for our analysis. 
 However, as $Q$ value (5435 keV) for $^{152}$Yb is higher,  corresponding $\gamma$ value for such nuclei are supposed to be small.

\begin{figure}[H]
\centering
\begin{tabular}{cc}
\includegraphics[width=0.4\textwidth]{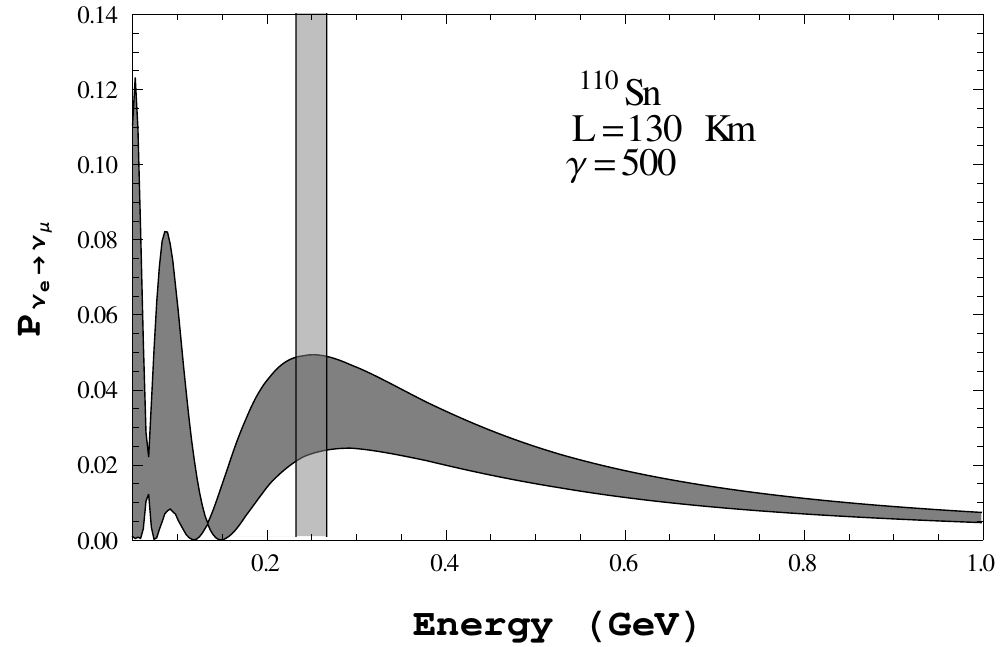}&
\includegraphics[width=0.4\textwidth]{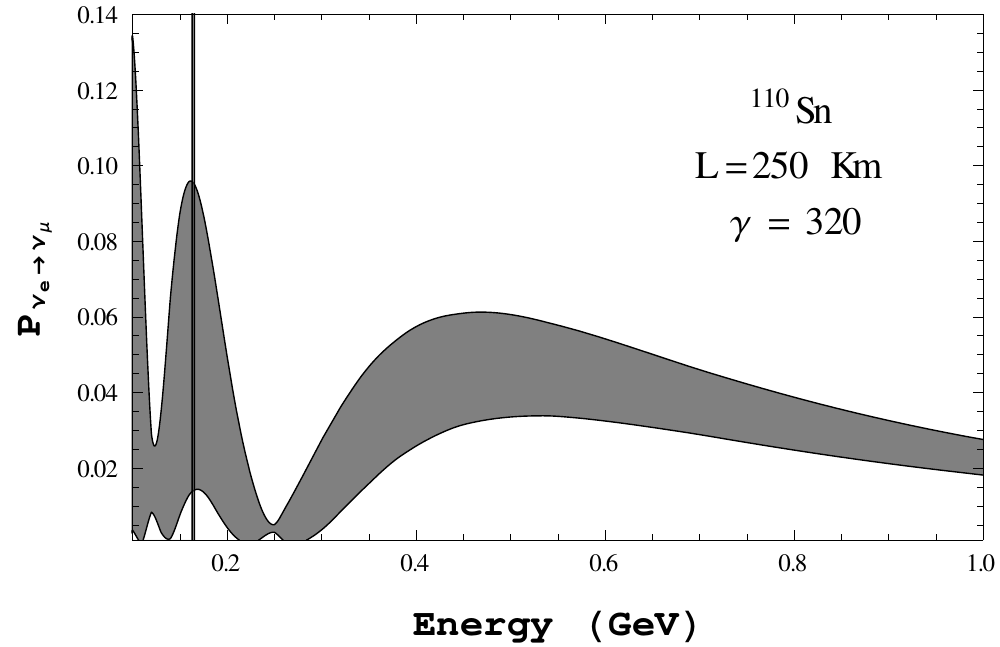}\\
\includegraphics[width=0.4\textwidth]{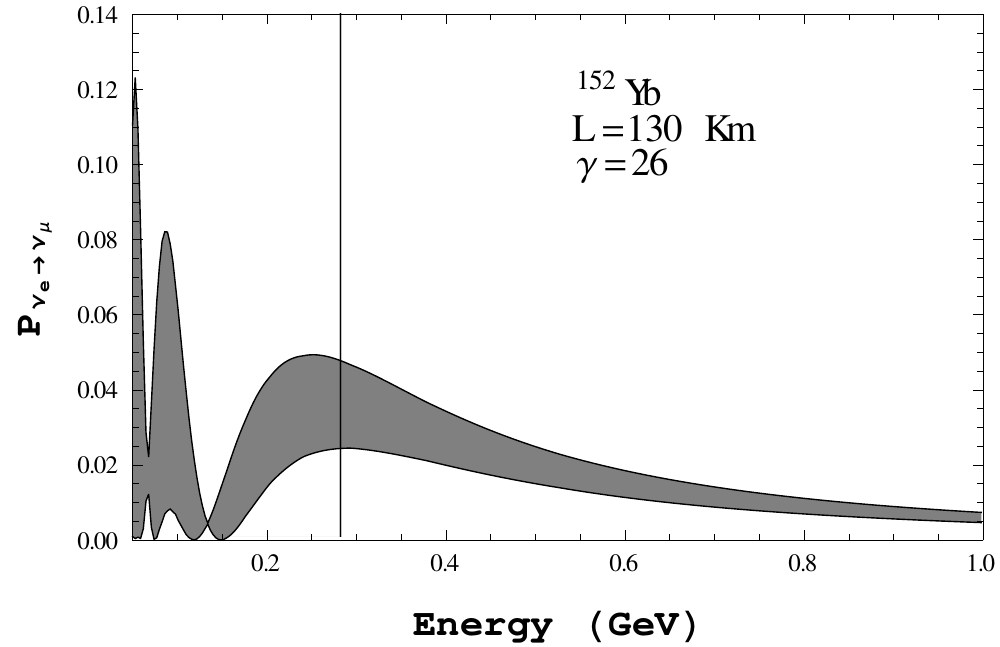}&
\includegraphics[width=0.4\textwidth]{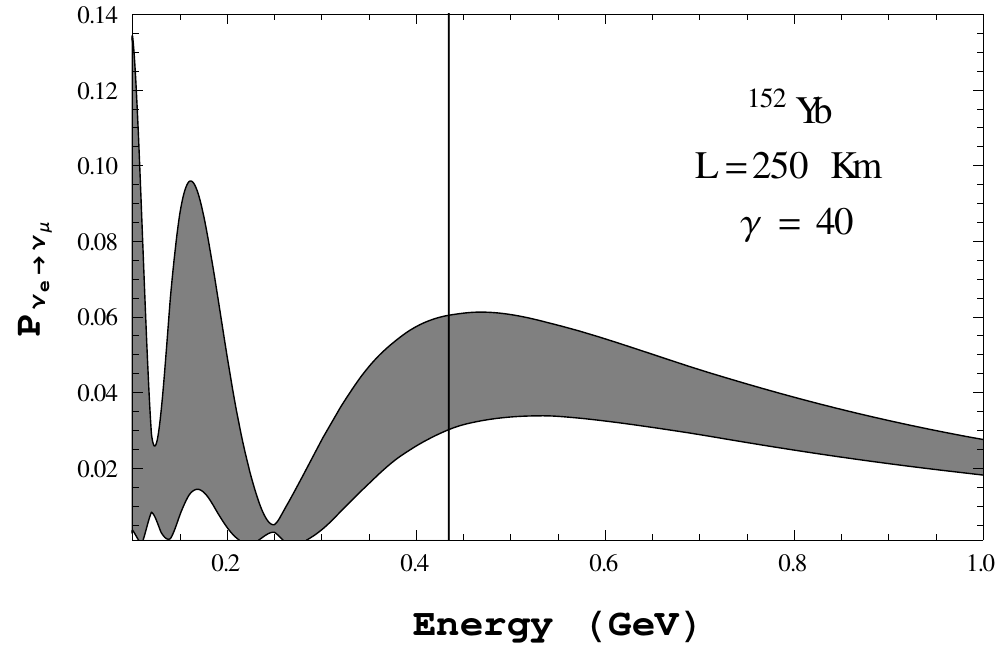}
\end{tabular}
\caption[] {{\small Probability $P(\nu_e\rightarrow \nu_{\mu})$ vs neutrino energy $E$ for 
two different nuclei $^{110}_{50}$Sn and $^{152}$Yb and the corresponding energy window satisfying equation (\ref{eq:ewindow}) for different $\gamma$ values.}} 
\label{fig:probmono}
\end{figure}

The  neutrino beam produced from electron capture process is boosted with certain boost factor, $\gamma$. The boosted neutrino beam produced from such process hits the detector at a baseline of length  $L$ at a radial distance $R$ from the beam axis and the energy, $E$ of this beam in rest frame of the detector , i.e., in laboratory frame is given by:

\begin{eqnarray}
E  (R)=\frac{Q}{\gamma}\bigg[1-\frac{\beta}{\sqrt{1+(R/L)^2}}\bigg]^{-1}\approx \frac{2\gamma Q}{1+(\gamma R/L)^2}
\label{eq:erest}
\end{eqnarray}
where R is the radial distance at the detector from the beam axis. At beam center, $R=0$. From the above equation (\ref{eq:erest}) the energy window considered for the analysis which is constrained by the size of the detector is given by:

\begin{equation}
\frac{2\gamma Q}{1+(\gamma R _{max}/L)^2} \leq E \leq 2\gamma Q
\label{eq:ewindow}
\end{equation}
 From equation (\ref{eq:ewindow}) we can see that once the baseline length L and $\gamma$ is fixed the energy window gets fixed.  
However, even considering  radius of the detector  $R_{max}=100 $ m the energy window is very small as can be seen from Figure \ref{fig:probmono}.

One can see from equation Eq.\eqref{eq:erest} that it is possible to tell precisely the energy from the $R$ value of the Cherenkov ring at the Cherenkov detector instead of measuring directly the neutrino energy. So there is scope to get good energy resolution by measuring position if the vertex resolution is good. The $\sigma(E)$ function corresponding to energy resolution function (as used in running GLoBES \cite{globes1})
 in terms of vertex measurement uncertainty $\sigma(R)$ can be written as:
\bea
\sigma(E)=-\frac{Q R \beta }{L^2 \left(1+\frac{R^2}{L^2}\right)^{3/2} \left(1-\frac{\beta }{\sqrt{1+\frac{R^2}{L^2}}}\right)^2 \gamma }\sigma(R)
\label{eq:ereso}
\eea
where $\beta$ is defined as
\bea
\beta=\frac{\sqrt{\gamma ^2-1}}{\gamma }
\label{eq:beta}
\eea
Vertex measurement uncertainty for electron (muon) identification at Super-K is around 30 (20) cm \cite{skvertex}. To estimate  the energy resolution using position measurement one may consider $\sigma(R) \sim 30 $ cm provided that the beam spreading $\frac{\sigma(R)}{L}$ is negligible ( lesser than
about $1 \mu$rad) \cite{rolinec} which is difficult to achieve experimentally.
 If we take into  account the beam divergence about 10 $\mu rad$  (which is 
almost one order larger than that considered in references \cite{ beam, divergence}), one may consider larger $\sigma(R) \sim 130 $ cm    particularly for baseline of 130 km. For baseline like 
250 km it would be more but we have considered same $\sigma(R)$ which means the beam divergence has been assumed to be lesser than about 5 $\mu rad$ for the analysis for  baseline with length 250 km. 

In this work GLoBES\cite{globes1}  has been used for doing the simulations.
In order to use this software, the radial binning is replaced by binning in energy and the bins are not equidistant.
If we divide $R^2_{max}$ into $k$ bins  the edges of the bins are given as:
\begin{eqnarray}
\label{eq:r}
R^2_i=R^2_{max}-(i-1)\Delta R^2 
\end{eqnarray}
with
\begin{eqnarray}
\label{eq:dr}
\Delta R^2=\frac{R^2_{max}}{k} 
\end{eqnarray}
We consider $R^2_i>R^2_{i+1}$ so that in GLoBES the respective energy bins are in the correct order as given below
\begin{eqnarray} 
\label{eq:ep}
E(R^2_i)<E(R^2_{i+1})
\end{eqnarray}
where 
$E$ is the neutrino energy in the lab frame.

The number of events per bin $i$ and channel $c$ (different channels mentioned later in this section)  is given by:
\begin{eqnarray}
\label{eq:event}
N_{event} \simeq \frac{N_{norm}}{L^2} \int_{E_i - \Delta E_i/2}^{E_i + \Delta E_i/2} dE' \int_0^\infty
dE \phi (E) P^c(L,E) \sigma^c(E)  \epsilon^c(E') R^c(E,E')
\end{eqnarray}
where $N_{norm}$ is the normalization factor for using GLoBES and is related to length of the baseline, area and energy binning related to 
flux , number of target nuclei per unit target mass and number of nuclei decaying.  
$\epsilon^c $ is the signal efficiency in the respective channel,   $P^c(L,E)$ is the neutrino oscillation probability in particular channel, $\sigma^c(E)$ is total cross section for particular flavor of neutrinos and particular interaction corresponding to particular channel. $R(E,E')$ is the energy resolution function of the detector where $E'$ is the reconstructed neutrino energy. $\phi (E)$ has been calculated from  the angular neutrino flux 
$\frac{dn}{d\Omega}(E)$ as defined below :
\bea
\label{eq:flux}
\frac{dn}{d\Omega}=\frac{N_{decays}}{4\pi}\left(\frac{E}{Q}\right)^2
\eea
where $N_{decays}$ is number of nuclei actually decaying per year. The detailed derivation of these expressions can be found in \cite{rolinec}. Considering equation (\ref{eq:ewindow})
and ( \ref{eq:flux}) one can see that with increase in $\gamma$ value the angular flux increases.

We have plotted numerically the probability $P(\nu_{e}\rightarrow \nu_{\mu})$ with respect to energy for two different baselines of length 130 km (CERN-Fr\'ejus) and 250 km  for
two different nuclei $^{110}_{50}$Sn and $^{152}$Yb. We have considered normal hierarchy in plotting Figure \ref{fig:probmono}. For all plots in Figure \ref{fig:probmono} ,  $\delta $ has been varied over its' entire allowed range (0 to 2 $\pi$) resulting in the shaded region in each plots 
showing the significant variation of probability at particular neutrino energies. Corresponding to each of
the nuclei (whose $Q$ values are fixed) we have considered appropriate $\gamma$ value  
so that the corresponding energy window (as mentioned in (\ref{eq:ewindow})) overlaps with
the shaded region near the oscillation peaks having significant variation of probability due to $\delta$ variation. In choosing $\gamma$, one also has to keep in mind that the neutrino energy is not too low as otherwise flux will be much lesser. The energy window has been shown by the shaded vertical strips. For our suitable choice of $\gamma$ value, the energy window  is larger for 130 km baseline and relatively smaller for 250 km baseline for both the nuclei. Also the energy window for $^{110}_{50}$Sn is larger than  $^{152}$Yb.

 For finding $\delta$ we shall prefer  the maximum variation of the probability with $\delta$ which will occur for neutrino energy satisfying $\frac{\Delta m ^2_{31}L}{4E} \approx (2n+1) \pi/2$ where $n$ is an integer. This has been shown in  Figure \ref{fig:probmono} in which the oscillation probability has been evaluated numerically considering the evolution of neutrino flavor states.  However, the energy also depends on the $Q$ value of the corresponding nuclei. So we have considered the case of two nuclei separately. As for example, for $^{110}_{50}$Sn nuclei, for baseline of length 130 km we have considered first oscillation maximum and for baseline of length 250 km we have considered second oscillation maximum where the variation of the probability due to $\delta $ is significant. For $^{152}$Yb, for both the baselines we have considered first oscillation peak.  
 In considering the suitable peak in the oscillation probability one has to keep in mind that the neutrino flux increases with $E^2$ 
as shown in Eq.\eqref{eq:flux} and so after doing the numerical analysis one can decide which 
energy out of various energies near various peaks are suitable. However, one also have to think about feasible boost factor $\gamma $.
For that reason although the first oscillation peak for $^{110}_{50}$Sn nuclei corresponds to higher energy with respect to 
2nd oscillation peak but as it also requires higher boost factor $\gamma $ around 900, we have considered the neutrino energy near the second
oscillation peak for $^{110}_{50}$Sn nuclei for baseline of length 250 km. Depending on the energy chosen near a peak one can appropriately choose  the boost factor $\gamma$ on which the neutrino energy window as shown in equation (\ref{eq:ewindow})  as well as  $\nu_e$ flux as shown in equation (\ref{eq:flux}) depend. This has been illustrated in Figure \ref{fig:probmono}.

\subsubsection{\bf Experimental set-ups and systematic errors}
For doing the analysis we choose four different set-ups:\
\newline
{\bf set-up(a)}: The length of the baseline is taken to be 130 km (CERN-Fr\'ejus baseline) and the boost factor $\gamma$ to be 500 for nuclei  $^{110}_{50}$Sn.
\newline
{\bf set-up(b)}: The length of the baseline is taken to be 250 km  and the boost factor $\gamma$ to be 320 for nuclei  $^{110}_{50}$Sn.
\newline
{\bf set-up(c)}: The length of the baseline is taken to be 130 km (CERN-Fr\'ejus baseline) and the boost factor $\gamma$ to be 26 for nuclei  $^{152}$Yb.
\newline
{\bf set-up(d)}: The length of the baseline is taken to be 250 km  and the boost factor $\gamma$ to be 40 for nuclei  $^{152}$Yb.

We consider a Water Cherenkov detector of fiducial mass 500 kt. Following \cite{signal}, the signal efficiency is considered to be 0.55 for  $\nu_\mu$ appearance channel. Background rejection factor coming from neutral current events is considered to be $10^{-4}$ for $\nu_\mu$ appearance channel. Signal error of $2.5\%$ and background error of $5\%$ has been considered.
For quasi-elastic $\nu_\mu$ appearance and $\nu_e$ disappearance we have followed signal efficiency and error as given in reference \cite{signal}.
We have considered the neutrino energy resolution as discussed earlier in (\ref{eq:ereso}) which can be obtained from vertex resolution after taking into account beam spreading.  The neutrino energy is known from
Eq.\eqref{eq:erest} and the energy width considered by us is obtained from Eq.\eqref{eq:ewindow} by considering the radius of the detector $R_{max}=100$m. We assume $10^{18}$ electron capture decays per year and the running time of  10 years for accumulating data. 

However, depending on the half life of Sn (4.11 hrs), the number of useful decays per effective  year ($10^{7}$ seconds ) considered are about
$0.608 \times 10^{18}$ with  boost factor ($\gamma = 500$) for    130 km baseline and $0.768 \times 10^{18}$ with boost factor ($\gamma = 320$) for 250 km baseline. Also, depending on the half life of $^{152}$Yb (3.04 seconds) and the boost factor $\gamma = 26$ or $\gamma = 40$, for baselines 130 km or 250 km respectively, the number of useful decays per effective  year ($10^{7}$ seconds ) considered are almost equal to the total number of nuclei i.e, $10^{18}$ as half life is  much smaller than  $^{110}_{50}$Sn. 
It is possible to achieve $\gamma$ about  480  at upgraded SPS facility at CERN \cite{sps,gam440,sps1} and $\gamma > 1000$ for LHC based design \cite{sps2}.
We have considered six energy bins keeping in mind the available energy window for different set-ups and the corresponding energy resolution in equation (\ref{eq:ereso}). In considering the energy resolution we have taken into account beam spreading. For that the the energy resolution considered by us is bad in comparison to the
energy resolution considered in reference \cite{rolinec} and we have to consider much lesser number of energy bins.  

As  the number of events corresponding to all set-ups  are quite large, any background due to atmospheric neutrinos are expected to be quite small and we have not considered such background in our analysis.

\subsubsection{\bf Possibilities of Discovery of $CP$ violation}

Here we discuss the discovery of $CP$ violation for four different experimental set-ups (a-d)
mentioned earlier for monoenergetic neutrino beam. 
In presenting our analysis we have followed the numerical procedure as discussed at the beginning of section  \ref{sec:sim}.  We have considered the true hierarchy as normal hierarchy. However, we have  considered the uncertainty in the hierarchy of neutrino masses in the test values as it is not known
at present. We have also considered the uncertainties in the other oscillation parameters as mentioned in table \ref{table:mix}.  For finding $CP$ violation we have fixed $\delta$(test) at $CP$ conserving $\delta$ values (0,$\pi$).  

\begin{figure}[H]
\centering 
\begin{tabular}{cc}
\includegraphics[width=0.4\textwidth]{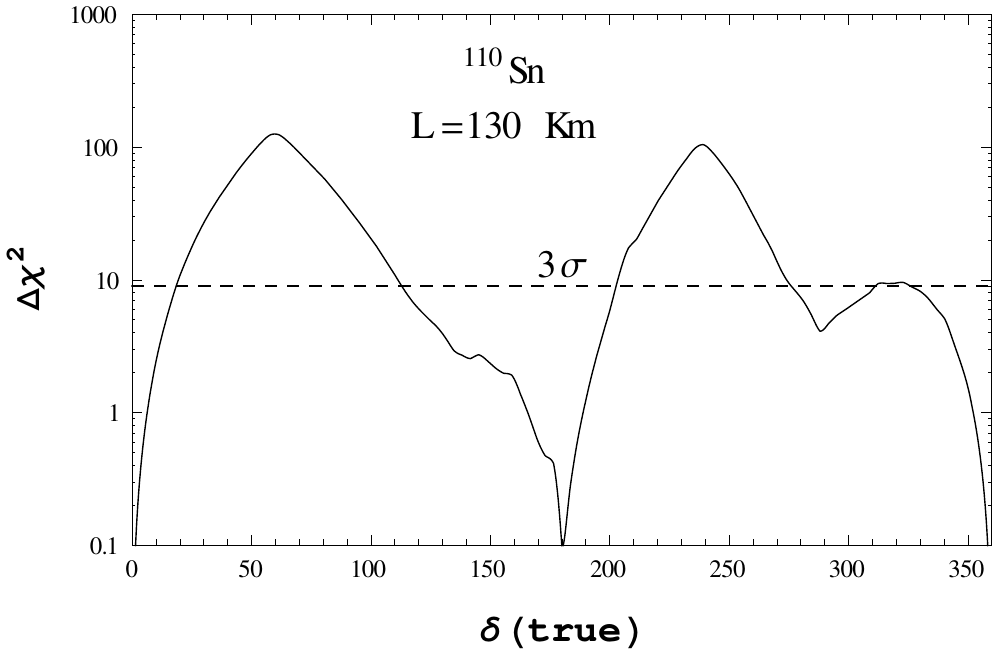}&
\includegraphics[width=0.4\textwidth]{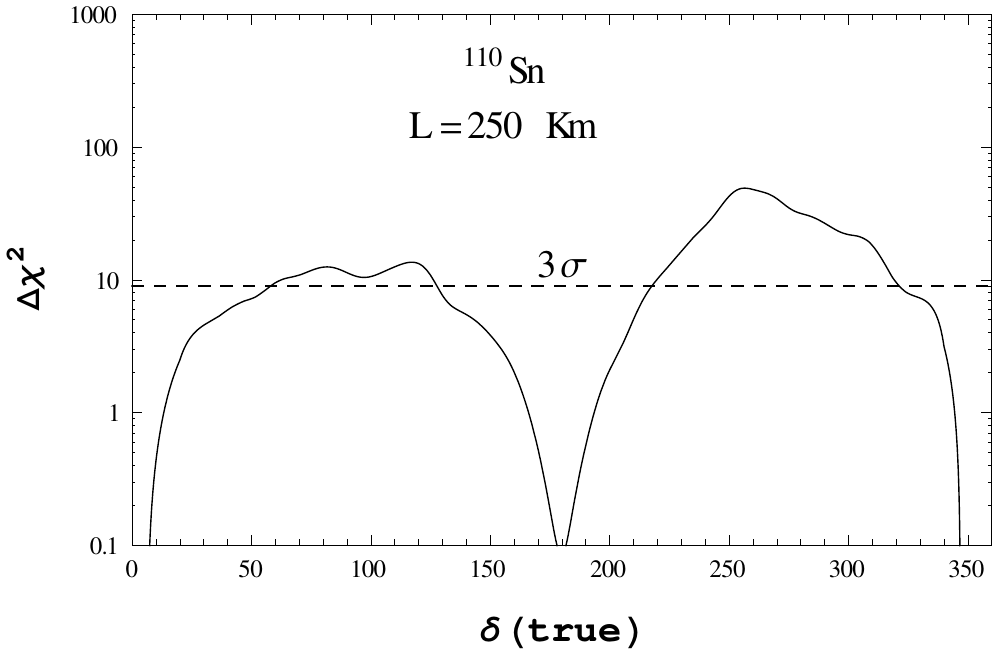}
\end{tabular}
\caption[] {{\small $\Delta \chi^2$ versus $\delta$ (true) for  two experimental set-ups 
(a) \& (b) with nuclei $^{110}_{50}$Sn for true normal hierarchy.}}
\label{fig2}
\end{figure}

\begin{figure}[H]
\centering
\begin{tabular}{cc}
\includegraphics[width=0.4\textwidth]{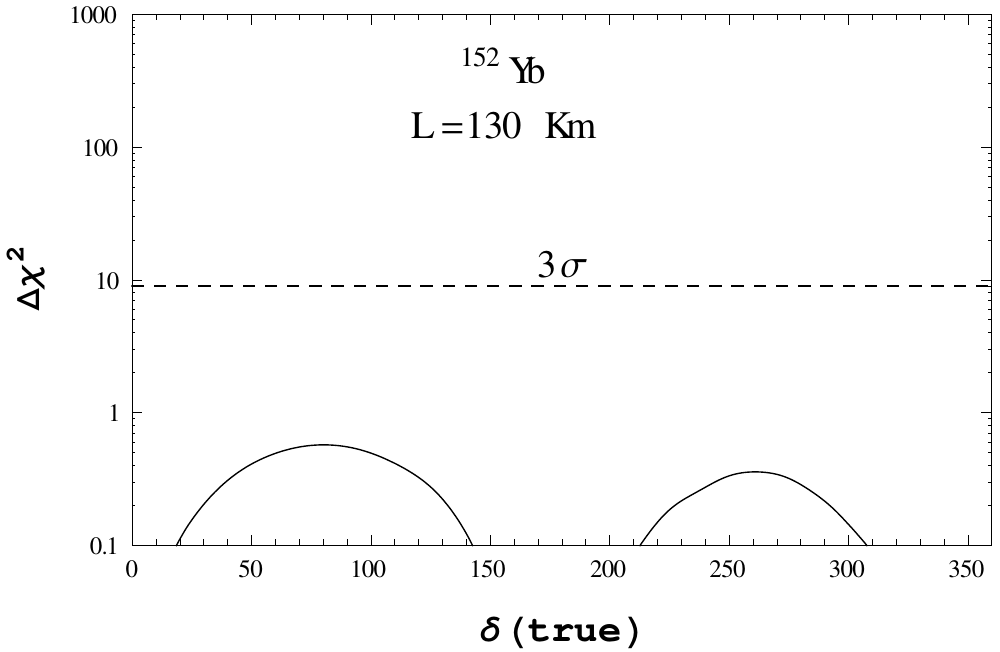}&
\includegraphics[width=0.4\textwidth]{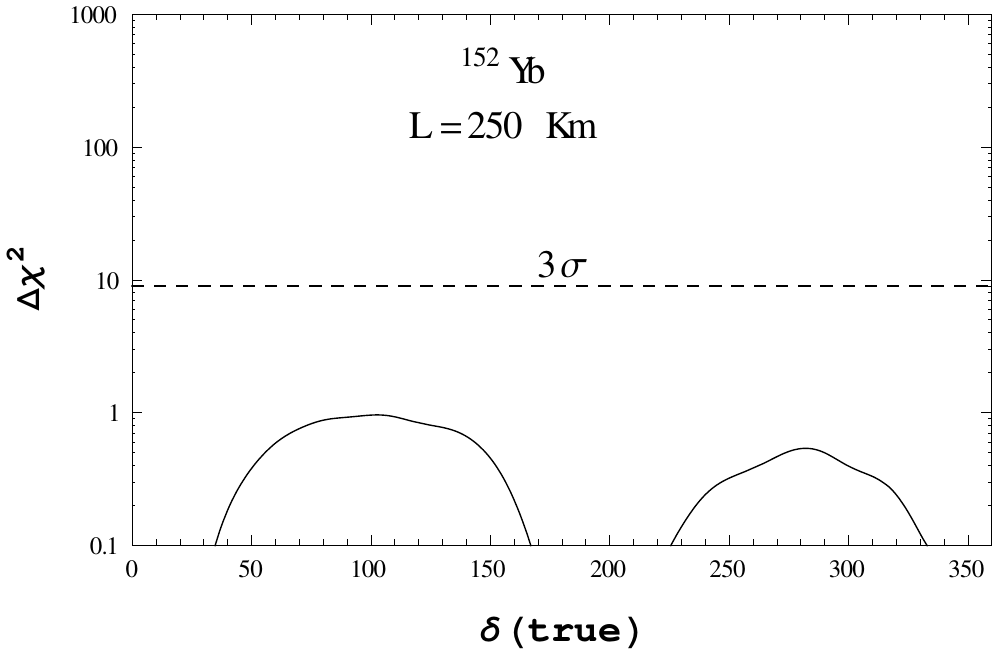}
\end{tabular}
\caption[] {{\small $\Delta \chi^2$ versus $\delta $(true) for  two  experimental set-ups 
(c) \& (d) with nuclei $^{152}$Yb for true normal hierarchy.}}
\label{fig3}
\end{figure}

In Figure \ref{fig2}, $\Delta \chi^2$ versus $\delta$ (true) has been plotted to show the discovery reach  of the $CP$ violation for two different set-ups - set-up(a) and set-up(b) for $^{110}_{50}$Sn nuclei.  We find that the discovery of $CP$ violation for set-up(a)  \&  (b)  
could be found with $F(\delta)$ for about 51\% and 49\% respectively of  the possible $\delta $ values
at $3\sigma$ confidence level.
In Figure \ref{fig3},  $\Delta \chi^2$ versus $\delta$ (true) has been plotted to show the discovery reach  of the $CP$ violation for two different set-ups - set-up(c) and set-up(d) for $^{152}$Yb nuclei.  We find that the discovery of $CP$ violation for set-up(c) and set-up (d) 
are not  good
and could not be found even at $1\sigma$ confidence level. For longer than 250 km baselines we have not
presented any plots for $CP$ violation discovery reach. It seems one of the basic problem
for longer baselines will be relatively bad energy resolution because we are trying to use vertex resolution for getting energy resolution but there is beam spreading and as such over longer baseline beam spreading will make the energy resolution poorer. 

In our analysis we have chosen neutrino energy near the oscillation peak (as shown in Figure \ref{fig:probmono}) which is more $\delta$ sensitive region. This consideration improves the $CP$ violation discovery reach.
However, in \cite{rolinec} (as shown in Figure 7 of that paper) $CP$ violation discovery reach has been shown to be  about 81\% of the possible $\delta $ values for their set-up II for 250 km baseline for the presently known
$\theta_{13}$ value. The discovery reach seems to be not so good  in our case although we have chosen appropriately the neutrino energy
where probaility of oscillation is $\delta$ sensitive. It is because
 in our analysis we have considered more realistic value of $\gamma$ (which is lower than that considered in \cite{rolinec} but could be achievable at present (keeping in mind the possible SPS upgrade at CERN). We have also taken into account the beam spreading in our present analysis in estimating  effective energy resolution
 and have taken only a few neutrino energy bins for the analysis unlike \cite{rolinec} as it is important to consider the size of the
 energy bins  larger in comparison to the level of energy resolution. Our results shows that the nuclei $^{110}_{50}$Sn is more suitable than the nuclei
$^{152}$Yb so far $CP$ violation discovery reach is concerned.

\section{Conclusion}
\label{sec:con}

We have studied the possible $CP$ violation discovery reach due to Dirac phase $\delta$ in the leptonic sector through neutrino oscillation experiments with superbeam, neutrino factory and mono-energetic neutrino beam from electron capture process.  
Particularly for superbeam and neutrino factory we have studied the NSI effect on the $CP$ violation discovery reach. 
However,  for mono-energetic neutrino beam we have considered shorter baseline as required technically for lower boost factor $\gamma $. For that NSI effect seems to be insignificant and hence we
have studied discovery of $CP$ violation due to $\delta $ without different NSI in this case.

For short baseline with CERN-Fr\'ejus  and T2HK set-up in case of superbeam for different NSI satisfying  model dependent bound  we find that there is
insignificant  effect due to real NSI on the the discovery reach of $CP$ violation due to $\delta$. For longer baseline with
CERN-Pyh\"asalmi set-up such effects are significant.  In case of neutrino factory, the baselines considered are slightly longer than those for superbeam. In this case, except $\ve_{\mu\mu}$ other NSIs have significant effect on $F(\delta)$. 

As $F(\delta)$ is never found to be 1, obviously for
some values of $\delta $,  $CP$ violation may not be observed at certain confidence level for the kind of detectors we have considered
(see Figure \ref{fig:delta1} ). 
Even if  one observes $CP$ violation due to $\delta $ in shorter baselines  there is a possibility of observing no $CP$ violation in
presence of real NSI. This
feature can be seen in Figure \ref{fig:delta2} for superbeam
in presence of real NSIs - $\ve_{ee}$, $\ve_{\tau\tau}$  and in Figure \ref{fig:cpfrac} in presence of real NSI  $\ve_{\tau \tau}$ for neutrino factory. For smaller values of NSI, the $CP$ violation discovery reach is much better for neutrino factory set-ups than those for superbeam set-ups. However, if NSIs like $\ve_{ee},\; \ve_{e\tau}$ and $\ve_{\tau\tau}$ are of significant strength $\gsim 0.1$ 
then $CP$ violation discovery reach at neutrino factory could be very bad in comparison to particularly T2HK set-up in superbeam.  One interesting feature is found from  Figure \ref{fig:delta2}. If  one does not observe $CP$ violation due to $\delta $ in shorter baselines (say CERN-Fr\'ejus set-up) there is still a possibility of observing  $CP$ violation
in longer baselines (see CERN-Pyh\"asalmi set-up with off-diagonal NSI like  $\ve_{e\mu}$ and $\ve_{e\tau}$) which could be the signal of NSI with significant strength. Using short and long baseline one could conclusively tell about $CP$ violation due to $\delta $ 
and about NSI under such situation.

NSI - $\ve_{e\mu}$, $\ve_{e\tau}$ and $\ve_{\mu\tau}$ could be complex. We have considered the corresponding phases 
$\phi_{e\mu}$, $\phi_{e\tau}$ and $\phi_{\mu\tau}$ respectively in the analysis of the discovery reach of total $CP$ violation 
for both in absence and in presence of $\delta $. Even in absence of $\delta $ one may observe $CP$ violation due to the
presence of NSI phases. The possibility of observing $CP$ violation under such scenario is relatively better in general for longer baselines
for both superbeam (see Figure \ref{fig:sbnsid0} ) and neutrino factory (see Figure \ref{fig:nsid0}) provided that absolute values of
NSIs are known. For $\ve_{e\mu}$ there is better $CP$ violation discovery reach in neutrino factory set-up. For $\ve_{e\tau}$ 
neutrino factory set-up with longer baseline is slightly better than CERN-Pyh\"aslmi set-up in superbeam at $3 \sigma$.
For $\ve_{\mu\tau}$ there is better $CP$ violation discovery reach for CERN-Pyh\"aslmi set-up in superbeam.

Assuming that there are two sources of $CP$ violation simultaneously existing in nature- one due to $\delta $ and the other due to say one of the NSI phases $\phi_{ij}$, the total $CP$ violation discovery reach has been shown as unshaded region in the $\phi_{ij} - \delta $ plane in Figure \ref{fig:sbnsiphase} for superbeam and Figure \ref{fig:phcol} for neutrino factory.   
From the Figures it is seen that  for NSI - $\ve_{e\mu}$ for  T2HK set-up there is more allowed region and as such better discovery reach is possible in the $\delta - \phi_{ij}$ parameter space particularly  in comparison to CERN-Pyh\"asalmi set-up in superbeam.  For NSI - $\ve_{e\tau}$ and $\ve_{\mu\tau}$, the discovery reach of total $CP$ violation is better for both longer and shorter baselines (like CERN-Pyh\"asalmi set-up and T2HK set-up). 
 As compared to
Figure \ref{fig:sbnsiphase} due to different superbeam set-ups, we find significantly better discovery reach of $CP$ violation for 
neutrino factory set-up for 730 km baseline in Figures \ref{fig:phcol}. Comparing Figures \ref{fig:phcol} and \ref{fig:sbnsiphase} we find that for $\phi_{e\tau}$   if we combine 
experimental data from superbeam set-ups and neutrino factory set-ups, there are scopes to improve the total $CP$ violation discovery 
reach further.
However, it would be difficult to disentangle  the observed $CP$ violation coming due to both  $\delta $ and NSI phase.
In presence of some ranges of off-diagonal NSI phase values $\phi_{e\mu}$ and $\phi_{\mu\tau}$ (see upper panel of Figure \ref{fig:sbnsiphase} for superbeam and Figure \ref{fig:phcol} for neutrino factory) there 
is possibility of discovering total $CP$ violation for any possible $\delta_{CP}$ value at $3 \sigma$.

The $CP$ violation due to $\delta $  could remain unobservable with present and near future experimental facilities
in superbeam and neutrino factory for certain range of values of $\delta $ (as for example for superbeam with T2HK set-up in Figure \ref{fig:delta1} for about $0^\circ \mbox{to} \;25^\circ$, $160^\circ \mbox{to} \;208^\circ$, $340^\circ \mbox{to} \; 360^\circ$ at 3 $\sigma $ confidence level). However, in presence of NSIs (with or without phases) the  $CP$ violation due to $\delta$ or
the total $CP$ violation due to $\delta$ and NSI phase could be observed 
even for such values of $\delta $ as can be seen in various Figures shown in Section III-A and B.

Basic strategy to find $CP$ violation in the leptonic sector in presence of  NSI for superbeam set-up may be to
consider  both shorter baseline
(say T2HK set-up)  as well as one  longer baseline (say CERN-Pyh\"asalmi set-up)  because of their complementary nature 
with respect to the discovery reach of $CP$ violation. If NSI values are smaller ($\lsim 0.1$) and real then 
the $CP$ violation discovery in neutrino factory set-up with MIND detector seems significantly  better than superbeam set-up.  
For complex NSIs, consideration of both superbeam set-up and neutrino factory set-up could give better $CP$ violation discovery reach.

We have  discussed only the possibility of discovering $CP$ violation. However, experimentally disentangling $CP$ violation
coming from $\delta$ present in PMNS matrix, the NSI phases and the absolute value of NSI could be very difficult. One may note that for longer baselines due to higher matter
density NSI effect could be more. Also the effect due to $\delta $ and that due to NSI on oscillation probability in particular channel varies differently with neutrino energy.  Only if the experimental data is available over certain range of neutrino energy from the shorter and longer  baseline experiments then only the multi-parameter fit
could help in disentangling  effects due to different unknown parameters.

For monoenergetic neutrino beam sources we have considered  two different 
nuclei - one   $\nu_e$ source is from electron capture  decays of $^{110}_{50}$Sn isotopes  and the other  $\nu_e$ source is from electron capture  decays of  isotopes $^{152}$Yb. 
For each case we have considered two baselines 130 km and 250 km. Among experimental set-ups 
(a-d)  in \ref{sec:mono} we find that the set-up (a) with  $^{110}_{50}$Sn isotope and  130 km baseline is found to be the most
suitable set-up for discovering $CP$ violation with $F(\delta)$ about 51 \% at $3 \sigma$ confidence level.

When one considers technical issues involved in the accelerator and running the ions through
vacuum tube, isotopes $^{152}$Yb  is better candidate than  $^{110}_{50}$Sn isotopes because of
much lesser half life.  $^{152}$Yb is also better because of the  dominant electron capture decay to one energy level. However, as can be seen from Figures \ref{fig2} and \ref{fig3},  the discovery reach of $CP$ violation is found to be better for  $^{110}_{50}$Sn isotopes.  Out of different baselines for $^{110}_{50}$Sn nuclei,  we find slightly better discovery reach for shorter baseline of 130 km with
$\gamma = 500$.

 Building up of such monoenergetic neutrino beam facilities  will require some technological development and the implementation of it might take some time \cite{Lind}. The existing CERN accelerator complex could be used to study such facility. However, the monoenergetic neutrino flux require a very large number of ions to be stored in the decay ring. It is difficult to control the beam at high intensities because of space charge detuning, intra beam scattering and vacuum loss.  With SPS upgrade it could be possible to accelerate the ions to $\gamma =480$ but accelerating above that seems difficult \cite{sps,gam440,sps1}. Depending on the half life
of $^{110}_{50}$Sn we have reduced the total number of useful decays of the ion per effective year from $10^{18}$ but the value considered is
still extreme because of the requirement of acceleration and storage of the partially charged ion. For improving this the vacuum conditions in
SPS would be required to be upgraded.  It requires more study on such beam facility. With technological improvement if it is possible to consider monoenergetic beam with $\gamma > 1000$
\cite{sps2}, then the $CP$ violation discovery reach will improve further than what has been
presented in this work.

\hspace*{\fill}

\noindent
\section*{Acknowledgments}  ZR  thanks University Grants Commission, Govt. of India for providing research fellowships. AD thanks Council of Scientific and Industrial Research, India for financial support through
Senior Research Fellowship (EMR No. 09/466(0125)/2010-EMR-I). We thank L. Whitehead  and Luca Agostino for providing  migration matrices for Liquid Argon detector and large scale water Cherenkov detector (as studied by MEMPHYS collaboration) respectively and for their other helpful communications. RA thanks R. Gandhi and S. K. Agarwalla for helpful discussion. We thank anonymous referee whose suggestions and comments have helped us immensely in improving the work.


\end{document}